\documentclass[journal]{IEEEtran}

\usepackage{cite}
\usepackage[numbers,sort&compress]{natbib}
\usepackage{amsmath,amssymb,amsfonts}
\usepackage{graphicx}
\usepackage{subfigure}
\usepackage{textcomp}
\usepackage{xcolor}
\usepackage{bm}

\usepackage{caption}

\usepackage{algpseudocode}  

\usepackage{algorithm}

\usepackage{amsthm}

\newtheorem{proposition}{Proposition}
\newtheorem{lemma}{Lemma}

\usepackage{stfloats}

\def\BibTeX{{\rm B\kern-.05em{\sc i\kern-.025em b}\kern-.08em
    T\kern-.1667em\lower.7ex\hbox{E}\kern-.125emX}}
    
\ifCLASSINFOpdf
\else
\fi

\begin{document}

\linespread {0.99}  
\addtolength{\parskip}{0.96pt} 

\setlength{\columnsep}{10pt}

\makeatletter
\renewcommand\normalsize{%
\@setfontsize\normalsize\@xpt\@xiipt
\abovedisplayskip 4\p@ \@plus2\p@ \@minus5\p@
\abovedisplayshortskip \z@ \@plus3\p@
\belowdisplayshortskip 6\p@ \@plus3\p@ \@minus3\p@
\belowdisplayskip \abovedisplayskip
\let\@listi\@listI}
\makeatother

\title{Synesthesia of Machines (SoM)-Enhanced Sub-THz ISAC Transmission for Air-Ground Network}


\author{Zonghui~Yang,~\IEEEmembership{Graduate~Student~Member,~IEEE}, Shijian~Gao,~\IEEEmembership{Member,~IEEE},\\
Xiang~Cheng,~\IEEEmembership{Fellow,~IEEE},
Liuqing~Yang,~\IEEEmembership{Fellow,~IEEE}
\thanks{Z.~Yang and X.~Cheng are with the State Key Laboratory of Photonics and Communications, School of Electronics, Peking University, Beijing 100871, China (e-mail: yzh22@stu.pku.edu.cn; xiangcheng@pku.edu.cn).}
\thanks{S.~Gao is with the Internet of Things Thrust, The Hong Kong University of Science and Technology (Guangzhou), Guangzhou 511400, China (e-mail: shijiangao@hkust-gz.edu.cn).}
\thanks{L.~Yang is with the Internet of Things Thrust \& Intelligent Transportation Thrust, The Hong Kong University of Science and Technology (Guangzhou) Guangzhou, China, and also with the Department of Electronic and Computer Engineering, Hong Kong University of Science and Technology, Hong Kong SAR, China (Email: lqyang@ust.hk).}
}
\maketitle

\markboth{IEEE TRANSACTIONS ON WIRELESS COMMUNICATIONS, 2025} %
{Shell \MakeLowercase{\textit{et al.}}: Bare Demo of IEEEtran.cls for IEEE Journals}

\vspace{-0.3cm}

\begin{abstract}

Integrated sensing and communication (ISAC) within sub-THz frequencies is crucial for future air-ground networks, but unique propagation characteristics and hardware limitations present challenges in optimizing ISAC performance while increasing operational latency. This paper introduces a multi-modal sensing fusion framework inspired by synesthesia of machine (SoM) to enhance sub-THz ISAC transmission. By exploiting inherent degrees of freedom in sub-THz hardware and channels, the framework optimizes the radio-frequency environment. Squint-aware beam management is developed to improve air-ground network adaptability, enabling three-dimensional dynamic ISAC links. Leveraging multi-modal information, the framework enhances ISAC performance and reduces latency. Visual data rapidly localizes users and targets, while a customized multi-modal learning algorithm optimizes the hybrid precoder. A new metric provides comprehensive performance evaluation, and extensive experiments demonstrate that the proposed scheme significantly improves ISAC efficiency.

\end{abstract}


\IEEEpeerreviewmaketitle

\vspace{-0.4cm}

\section{Introduction}


\IEEEPARstart{T}HE air-ground network is a foundational infrastructure for networked intelligence and low-altitude economy, addressing safety and efficiency needs in intelligent transportation, autonomous logistics, and other next-generation applications \cite{overview_LA_1}. These networks require high-speed data transmission and precise sensing capabilities \cite{cheng2022integrated}. Base stations (BS) must ensure the quality of services for communication users, such as mobile ground vehicles, while localizing low-altitude unmanned aerial vehicles (UAVs). To meet these dual demands, 6G has positioned integrated sensing and communication (ISAC) as a crucial enabling technology \cite{overview_5GA}. Achieving data rates up to 100 Gbps and sub-millimeter-level positioning accuracy necessitates designing ISAC transmission at sub-THz frequencies \cite{overview_THz_ISAC}. As the Internet of intelligent devices grows, future networks will incorporate non-radio frequency (RF) sensors, such as cameras \cite{overview_Som, overview_vision}, requiring BSs to integrate multi-modal sensing with sub-THz RF communications.

Designing efficient ISAC transmission in sub-THz systems faces three principal challenges. Firstly, the unique propagation characteristics of sub-THz systems hinder ISAC performance optimization and increase operational time overhead. Secondly, the three-dimensional and dynamic nature of air-ground networks introduces complex trade-offs in ISAC transmission. Thirdly, integrating heterogeneous sensing modalities remains an unresolved obstacle to enhancing ISAC capabilities. Addressing these challenges is crucial for developing robust and efficient air-ground networks that can meet the stringent demands of future applications.

Specifically, in sub-THz channels, beam squint effects can significantly degrade array gain, impairing dual-functional performance \cite{how_many_squint, beam_squint_estimation}. To counteract this, true time delay (TTD) has been incorporated into hybrid precoders, enabling frequency-dependent phase compensation for cross-subcarrier beam alignment \cite{DPP, DPP_constrained_TTD}. 
For sensing applications, \cite{CBS} proactively controlled squint patterns through TTDs to achieve frequency-aware target positioning.
Nevertheless, current sub-THz systems lack a unified design that optimally balances both functions. 
\cite{correlation_RIS_learning} explored the degree of freedom (DoF) of sub-THz channels through reconfigurable intelligent surfaces, expanding ISAC Pareto boundaries, but this method is primarily suited for narrowband systems and exacerbates beam squint in sub-THz. 
Beyond channel limitations, sub-THz hardware introduces operational complexity, particularly for channel state information (CSI) acquisition and precoder optimization.
\cite{doubly_selective, THz_CE_JSAC} accelerated sub-THz channel estimation by exploiting sparsity and low-rank nature, but neglected sensing-induced interference. \cite{ISAC_beamforming_conghui, ISAC_THz_opt} employed block coordinate descent (BCD) methods for efficient hybrid precoding, yet incurred performance degradation in sub-THz configurations.

Meanwhile, the three-dimensional topology and dynamic characteristics of air-ground networks pose critical difficulties for sub-THz ISAC transmission.
These challenges arise from the need to accommodate complex spatial configurations and movement patterns, which introduce intricate trade-offs in signal processing and transmission strategies.
To achieve comprehensive coverage across both horizontal and elevation directions, deploying uniform planar arrays (UPA) becomes essential \cite{UPA_JSAC, beam_training_UPA}. Unfortunately, this spatial expansion introduces more flexible beam dispersion in sub-THz compared to linear arrays. Hence, existing methodologies designed for linear arrays, such as \cite{beam_tracking_tanjingbo, beam_tracking_shixu}, cannot be directly extended to such three-dimensional scenarios.
Recent advances like \cite{UPA_TTD_1} embedded two-dimensional TTD arrays for UPA to alleviate beam squint, enabling three-dimensional communication links. Nonetheless, the UPA-based three-dimensional sensing under beam squint remains unexplored. This gap necessitates customized beam squint manipulation for UPA to establish a dual-functional trade-off.
Furthermore, both ground users and low-altitude targets exhibit high mobility.
Although \cite{dynamic_MP} developed predictive precoding for vehicular mobility, they assume ideal user-target alignment, a questionable premise. \cite{TCOM} deployed ISAC in doubly dynamic networks yet neglected low-altitude links, failing to support air-ground networks.

While leveraging multi-modal sensing data from BS shows promise in addressing these challenges, aligning heterogeneous data and integrating them with ISAC introduces a new set of complexities. Inspired by the concept of machine synesthesia (SoM) \cite{overview_Som}, out-of-band information has been leveraged for transceiver optimization through judicious network design, bringing potential for efficiency improvement \cite{Lidar_channel_model, MMFF_Som, vision_channel_estimation, TCOM, FR}. For example, \cite{MMFF_Som} utilized computer vision for rapid user detection and beam prediction, while \cite{vision_channel_estimation} derived channel covariance matrices based on visual and positioning knowledge.
However, RF sensing remains underexplored in these schemes. To advance ISAC,
\cite{TCOM} dynamically adjusted mmWave hybrid precoders using CSI-GPS fusion, yet encounters prohibitive sub-THz complexity. Furthermore, positioning priors alone cannot fully characterize channel states, limiting the performance. \cite{FR} developed a SoM-driven analog precoder to extend sub-THz ISAC performance boundaries, but neglects the time overhead of prior acquisition and precoder design, rendering it impractical for real-world deployment.

To address the aforementioned challenges, we propose a novel paradigm of SoM-enhanced sub-THz ISAC transmission for air-ground networks that simultaneously improves dual-functional performance while minimizing time overhead. This innovative methodology leverages the intrinsic DoFs present in sub-THz hardware and channels, and synergistically fuses visual data with RF measurements. To mitigate impairments associated with sub-THz channels, we first establish the critical dependence of ISAC performance on the correlation between communication and sensing channels. This allows us to proactively establish equivalent channels for ISAC enhancement without the need to modify existing hardware.
For three-dimensional air-ground deployments, we initiate the process by employing visual sensing for real-time user and target detection. Subsequently, we develop a squint-aware cross-pattern beam tracking (SA-CP-BT) mechanism for UPA, enabling rapid and precise angle refinement. Building on these advancements, we design a hybrid precoder through multi-modal data fusion via a dedicated neural network that utilizes a customized loss function. To comprehensively evaluate the practical performance of our approach, we introduce the concept of ISAC efficiency, a holistic metric that balances performance gains against the corresponding time overhead. Extensive simulations validate the superiority of the proposed framework, demonstrating its potential to enhance the effectiveness of ISAC in future air-ground networks.

The key contributions are summarized as follows:
\begin{itemize}
    \item The intrinsic DoF in sub-THz channels is leveraged to optimize ISAC performance by establishing a key relationship between communication-sensing channel correlation and system effectiveness. This approach uses existing hardware DoF to accelerate prior information acquisition and extend ISAC capabilities.
    \item A squint-aware transmission scheme for UPA is created to establish adaptable three-dimensional ISAC links within dynamic air-ground networks. ISAC efficiency is introduced as a comprehensive metric to evaluate and measure system performance.
    \item An innovative ISAC transmission paradigm integrates multi-modal sensing data, including RF measurements and visual information. This method enhances prior data acquisition and precoder design, significantly reducing time overhead while ensuring robust ISAC performance.
\end{itemize}

\begin{figure}[t]
  \vspace{-0.2cm}
  \centering
  \includegraphics[width=0.94\linewidth]{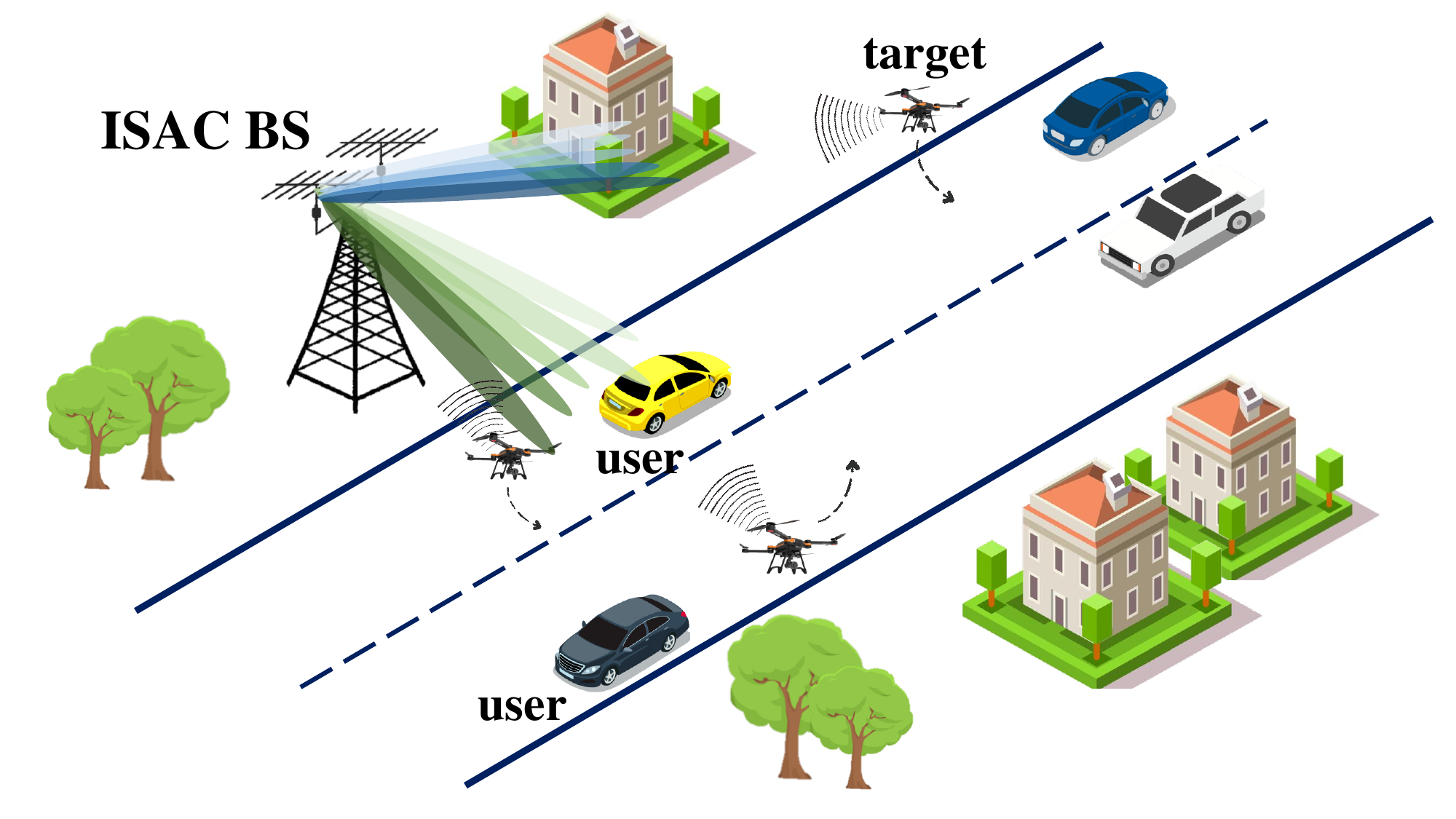}
  \vspace{-0.2cm}
  \caption{A diagram of ISAC systems in air-ground network.}
  \label{fig:system model}
  \vspace{-0.55cm}
\end{figure}

\newcommand{\RNum}[1]{\uppercase\expandafter{\romannumeral #1\relax}}

The remainder of this work is structured as follows. Section \RNum{2} introduces the system model. Section \RNum{3} exhibits the proposed SoM-enhanced prior information estimation scheme, and Section \RNum{4} explains the proposed SoM-enhanced ISAC precoding approach for sub-THz systems. Section \RNum{5} presents our experimental results, and Section \RNum{6} concludes our work.

\textit{Notation}: $a$, $\bm a$ and $\bm A$ represent a scalar, a vector and a matrix. $(\cdot)^{\text{T}}$, $(\cdot)^{\text{H}}$, $\left\|\cdot\right\|_2$, $\left\|\cdot\right\|_{\text{F}}$ denote the transpose, the conjugate transpose, 2-norm and Frobenius-norm. $\odot$ and $\otimes$ denote the Hadamard product and the Kronecker product. $\text{diag}(\bm a)$ and $\text{diag}(\bm A)$ denote the vector diagonal matrixization and extracting diagonal elements from a matrix. $\mathcal{U}[a,b]$ denotes the uniform distribution between $a$ and $b$. $\mathcal{CN}(m,\sigma^2)$ represents the complex Gaussian distribution with mean being $m$ and covariance being $\sigma^2$. $\mathbb{E}(\cdot)$ denotes the expectation. $\bigcup$ denotes the union of sets and $\backslash$ denotes the difference set.
$\mathbb{R}$ and $\mathbb{C}$ denote the sets of real numbers and complex numbers. $\text{Re}$ and $\angle$ denote the real part and the phase of the complex numbers.

\section{System Model}

We consider a wideband multi-antenna ISAC BS that serves $U$ single-antenna ground users while simultaneously sensing $K$ nearby low-altitude targets, as illustrated in Fig.~\ref{fig:system model}.
The ISAC signal is transmitted using orthogonal frequency division multiplexing (OFDM) with $M$ subcarriers. The frequency of the $m$-th subcarrier is $f_{m} = f_{\text{c}} + \frac{B}{M} \left( m - \frac{M+1}{2} \right)$, with $f_{\text{c}}$ being the carrier frequency and $B$ being the bandwidth.
The BS is equipped with a UPA consisting of $N_{t}=N_{t_{h}}\times N_{t_{v}}$ antennas. Additionally, a co-located UPA with $N_{r}=N_{r_{h}}\times N_{r_{v}}$ antennas is used for echo reception.

Furthermore, RGB-Depth (RGB-D) cameras are placed at the BS to oversee the surroundings. The field of view (FOV) of cameras is denoted as $fov$ (in radians). $N_{w}$ and $N_{h}$ indicate the number of pixels horizontally and vertically.

\begin{figure*}[t]
  \vspace{-0.0cm}
  \centering
  \includegraphics[width=0.86\linewidth]{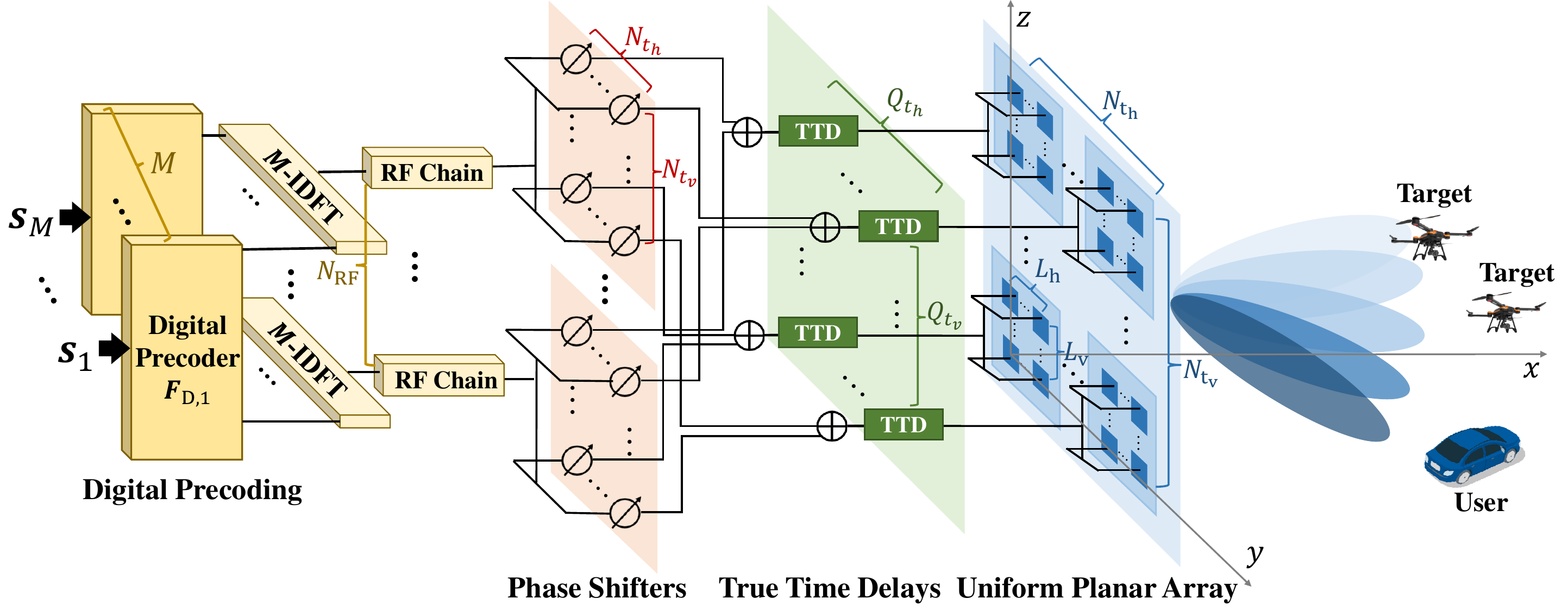}
  \vspace{-0.0cm}
  \captionsetup{font=small}
  \caption{A diagram of hybrid precoding structure in sub-THz systems.}
  \label{fig:precoding structure}
  \vspace{-0.35cm}
\end{figure*}

\subsection{Hybrid Precoder}

As illustrated in Fig.~\ref{fig:precoding structure}, a typical hybrid architecture is employed in the sub-THz system, where TTDs are placed between the PSs and the antenna array. This configuration facilitates the frequency-dependent adjustment of phase shifts to alleviate beam squint effects. To start with, data across $M$ subcarriers are first processed by frequency-dependent digital precoders, followed by $N_{\text{RF}}$ $M$-point inverse fast Fourier transforms (IFFT).
Then $N_{\text{RF}}$ RF chains, each connected to $N_{t}$ PSs, are deployed. After passing through these PSs, the signals are combined by $Q_{t}=Q_{t_{h}}\times Q_{t_{v}}$ TTD units. Each TTD is associated with $L_{t}=L_{h}\times L_{v}$ PSs for each RF chain, where $L_{h}=\frac{N_{t_{h}}}{Q_{t_{h}}}$ and $L_{v}=\frac{N_{t_{v}}}{Q_{t_{v}}}$. Additionally, the UPA is partitioned into $Q_{t}=Q_{t_{h}}\times Q_{t_{v}}$ sub-arrays, each connected to one TTD at the corresponding location.
Thus, the transmitted signal at the $m$-th subcarrier can be written as
\vspace{-0.15cm}
\begin{equation}
    \bm x_{m} = \bm F_{\text{T},m} \bm F_{\text{P}} \bm F_{\text{D},m} \bm s_{m}.
    \vspace{-0.15cm}
    \label{equ:x_m}
\end{equation}
In Eq. (\ref{equ:x_m}), $\bm s_{m}=[s_{1,m}, \cdots, s_{U,m}]^{\text{T}}\sim\mathcal{CN}(0, \bm I_{U})$ represents the data symbols for $U$ users at the $m$-th subcarrier. $\bm F_{\text{D},m} \in \mathbb{C}^{N_{\text{RF}} \times U}$ is the digital precoder at the $m$-th subcarrier. $\bm F_{\text{P}} \in \mathbb{C}^{N_{t} \times N_{\text{RF}}}$ denotes the frequency-independent precoder executed by the PS network, with $\vert\bm F_{\text{P}} [i,j]\vert = 1$. The frequency-dependent phase shifts introduced by the TTDs can be represented as 
\vspace{-0.15cm}
\begin{equation}
    \bm F_{\text{T},m} = \text{diag} \left[ \text{vec} \left( e^{j2\pi f_{m} (\bm T \otimes \bm 1^{L_{h}\times L_{v}})} \right) \right],
    \vspace{-0.15cm}
\end{equation}
where $\bm T \in [0, t_{\text{max}}]^{Q_{t_{h}} \times Q_{t_{v}}}$ is associated with the TTD's controllable delays, and $\bm 1^{L_{h} \times L_{v}}$ is an all-one matrix. $t_{\text{max}}$ indicates the maximum achievable time delay determined by the hardware.

\vspace{-0.1cm}
\subsection{Communication Model}

Similar to \cite{com_channel}, the communication channel between the BS and the $u$-th user at the $m$-th subcarrier is modeled as
\vspace{-0.2cm}
\begin{align}
    \bm h_{u,m}&=\beta_{u,m} e^{-j\frac{2\pi f_{m}d_{\text{c},u}}{c_{v}}}\bm a_{\text{t}}(\theta_{\text{c},u}, \phi_{\text{c},u}, f_{m}) \\
    &+ \frac{1}{\sqrt{P_{u}}K_{f}}\!\sum_{p=1}^{P_{u}}\beta_{u,p,m} e^{-j\frac{2\pi f_{m}d_{\text{c},u,p}}{c_{v}}}\!\bm a_{\text{t}}(\theta_{\text{c},u,p}, \phi_{\text{c},u,p}, f_{m}). \nonumber
\end{align}
For user-$u$, $\theta_{\text{c},u}\in [0,\pi]$ and $\phi_{\text{c},u}\in [-\pi/2,\pi/2]$ are the elevation and the azimuth angles of the line-of-sight (LoS) path. $\beta_{u,m}=\frac{c_{v}}{4\pi f_{m}d_{\text{c},u}}e^{j\upsilon_{\text{c},u}}$, $\beta_{u,p,m}=\frac{c_{v}}{4\pi f_{m}d_{\text{c},u,p}}e^{j\upsilon_{\text{c},u,p}}$ denote the path gain of the LoS path and non-line-of-sight (nLoS) ones. Here, $d_{c,u}$ and $d_{\text{c},u,p}$ represent the propagation distances and $\upsilon_{\text{c},u}$ and $\upsilon_{\text{c},u,p}$ are random noise phase, modeled as a uniformly distributed variable.
$\bm a_{t}(\theta_{\text{c},u}, \phi_{\text{c},u}, f_{m})=\bm a_{h}(\theta_{\text{c},u}, \phi_{\text{c},u}, f_{m})  \otimes \bm a_{v}(\theta_{\text{c},u}, f_{m})\in \mathbb{C}^{N_{t}\times 1}$ denotes the steering vector at $f_{m}$ from the transmitter towards $\theta_{\text{c}}$ and $\phi_{\text{c}}$,
with $[\bm a_{h}(\theta, \phi,  f_{m})]_{n_{t_{h}}}=\frac{1}{\sqrt{N_{t_{h}}}} e^{j\pi \frac{f_{m}}{f_{\text{c}}}\sin{\phi}\sin{\theta}(n_{t_{h}}-1)}$, $[\bm a_{v}(\theta, f_{m})]_{n_{t_{v}}}=\frac{1}{\sqrt{N_{t_{v}}}} e^{j\pi \frac{f_{m}}{f_{\text{c}}}\cos{\theta}(n_{t_{v}}-1)}$.
Based on the above setup, the received signal at the $u$-th user at the $m$-th subcarrier is represented as
\vspace{-0.15cm}
\begin{equation}
y_{\text{c},u,m}=\bm h_{u,m}^{\text{H}}\bm F_{\text{T},m}\bm F_{\text{P}}\bm F_{\text{D},m}\bm s_{m}+z_{\text{c},u,m},
\vspace{-0.1cm}
\end{equation}
where $z_{\text{c},u,m}$ is additive white Gaussian noise with power spectrum density $n_{\text{c},0}$ (W/Hz). Taking into account inter-user interference, the corresponding spectrum efficiency (SE) becomes
\vspace{-0.1cm}
\begin{equation}
    \mathcal{R}_{u,m}\!=\!\frac{B}{M}\!\log_{2}\!\left(\!1\!+\! \frac{\vert \bm h_{u,m}^{\text{H}}\bm F_{m}[:,u]\vert^{2}}{\sum_{i\neq u}\!\vert \bm h_{u,m}^{\text{H}}\bm F_{m}[:,i]\vert^{2}\!+\!\frac{B}{M}n_{\text{c},0}} \right),
    \vspace{-0.1cm}
\end{equation}
where $\bm F_{m}=\bm F_{\text{T},m}\bm F_{\text{P}}\bm F_{\text{D},m}$.
Consequently, the overall spectrum efficiency (SE) is computed as
\vspace{-0.15cm}
\begin{equation}
    \mathcal{R}=\sum_{m=1}^{M}\sum_{u=1}^{U}\mathcal{R}_{u,m}.
    \vspace{-0.15cm}
\end{equation}
$\mathcal{R}$ will be optimized by acquiring prior channel information and then utilizing an appropriate hybrid precoding strategy.

\vspace{-0.2cm}
\subsection{Sensing Model}
According to \cite{sensing_channel}, the target response matrix at the $m$-th subcarrier can be expressed as
\vspace{-0.1cm}
\begin{equation}
    \bm G_{m}=\bm A_{r,m} \bm\Sigma_{m}\bm A_{t,m}^{\text{H}},
\label{equ:sensing channel}
\vspace{-0.2cm}
\end{equation}
where $\bm A_{t,m}=[\bm a_{\text{t}}(\theta_{\text{s},1}, \phi_{\text{s},1},f_{m}),\cdots,\bm a_{\text{t}}(\theta_{\text{s},K}, \phi_{\text{s},K},f_{m})]$, and $\Sigma_{m}=\text{diag}(\alpha_{1,m}, \cdots,\alpha_{K,m})$. $\alpha_{k,m}=\sqrt{\frac{c_{v}^{2}\sigma_{\text{RCS},k}}{(4\pi)^{3}f_{m}^{2}d_{\text{s},k}^{4}}}e^{j\upsilon_{\text{s},k}}$ denotes the complex reflection coefficient of the $k$-th target, with $\sigma_{\text{RCS},k}$ and $d_{\text{s},k}$ being the RCS and the distance of the $k$-th target respectively, and $\upsilon_{\text{s},k}\sim \mathcal{U}[0,2\pi]$.
$\bm \theta_{\text{s}}=[\theta_{\text{s},1}, \cdots, \theta_{\text{s},K}]$ and $\bm \phi_{\text{s}}=[\phi_{\text{s},1}, \cdots, \phi_{\text{s},K}]$ denote the elevation and azimuth angles of the $K$ targets respectively. 
Then the echo at the $m$-th subcarrier at BS can be written as
\vspace{-0.15cm}
\begin{equation}
    \bm y_{\text{s},m}= \bm G_{m} \bm F_{\text{T},m} \bm F_{\text{P}}\bm F_{\text{D},m} \bm s_{m}+\bm z_{\text{s},m},
    \vspace{-0.15cm}
\end{equation}
where $\bm z_{\text{s},m}$ denotes the ISAC receiver white Gaussian noise with power spectral density $n_{\text{s},0}$ (W/Hz). To ensure that the noise follows a complex Gaussian distribution, we consider whitening the received noise before target sensing.
Accordingly, the Cramér-Rao bound (CRB) associated with the unbiased angle estimation can be derived as
\vspace{-0.15cm}
\begin{equation}
    \mathcal{CRB}(\bm \theta_{\text{s}}, \bm \phi_{\text{s}})=\text{Tr}\left[\left(\sum_{m=1}^{M}\mathcal{J}_{m}\right)^{-1}\right].
    \vspace{-0.1cm}
\end{equation}
$\mathcal{J}_{m}$ is the Fisher information matrix with respect to $\bm \theta_{\text{s}}$ and $\bm \phi_{\text{s}}$ given by
\vspace{-0.1cm}
\begin{equation}
\mathcal{J}_{m}=\frac{2}{Bn_{\text{s},0}}\text{Re}
\left[ 
\begin{matrix}
\mathcal{J}_{m}(\bm \theta_{\text{s}}, \bm \theta_{\text{s}}), &  \mathcal{J}_{m}(\bm \theta_{\text{s}}, \bm \phi_{\text{s}})   \\
\mathcal{J}_{m}^{\text{H}}(\bm \theta_{\text{s}}, \bm \phi_{\text{s}}), & \mathcal{J}_{m}(\bm \phi_{\text{s}}, \bm \phi_{\text{s}}) \\
\end{matrix}
\right].
\vspace{-0.15cm}
\end{equation}
Each entry of $\mathcal{J}_m$ can be computed as
\vspace{-0.1cm}
\begin{equation}
\begin{aligned}
    \mathcal{J}_{m}(\bm \theta_{\text{s}}, \bm \phi_{\text{s}})&\!=\!(\dot{\bm A}_{r,\theta,m}^{\text{H}} \dot{\bm A}_{r,\phi,m})\!\odot\! (\bm \Sigma_{m}^{*}\bm A_{t,m}^{\text{H}}\bm R_{x,m}^{*}\bm A_{t,m}\bm \Sigma_{m} ) \\
    &+\! (\dot{\bm A}_{r,\theta,m}^{\text{H}} \bm A_{r,m})\!\odot\! (\bm \Sigma_{m}^{*}\bm A_{t,m}^{\text{H}}\bm R_{x,m}^{*}\dot{\bm A}_{t,\phi,m}\bm \Sigma_{m} ) \\
    &+\!(\bm A_{r,m}^{\text{H}} \dot{\bm A}_{r,\phi,m})\!\odot\! (\bm \Sigma_{m}^{*} \dot{\bm A}_{t,\theta,m}^{\text{H}}\bm R_{x,m}^{*}\bm A_{t,m}\bm \Sigma_{m} ) \\
    &+\!(\bm A_{r,m}^{\text{H}} \!\bm A_{r,m})\!\odot\! (\bm \Sigma_{m}^{*}\dot{\bm A}_{t,\theta,m}^{\text{H}}\!\bm R_{x,m}^{*}\!\dot{\bm A}_{t,\phi,m}\bm \Sigma_{m} ),  \nonumber
\end{aligned}
\end{equation}
with $\bm R_{x,m}=\mathbb{E}\left[\bm x_{m}\bm x_{m}^{\text{H}}\right]$, $\dot{\bm A}_{t,\theta,m}[:,k]=\frac{\partial \bm a_{t}(\theta_{\text{s},k}\phi_{\text{s},k},f_{m})}{\partial \theta_{\text{s},k}}$ and $\dot{\bm A}_{t,\phi,m}[:,k]=\frac{\partial \bm a_{t}(\theta_{\text{s},k},\phi_{\text{s},k},f_{m})}{\partial \phi_{\text{s},k}} ]$.


In conventional optimization frameworks, $(\bm \theta_{\text{s}}, \bm \phi_{\text{s}})$ is typically obtained through preamble-based beam scanning procedures \cite{crb_opt}, which introduces significant overhead. The proposed method instead acquires this prior information through multi-modal sensing at the BS, establishing the foundation for efficient precoder design.

\subsection{ISAC Performance Enhancer: C-S Channel Correlation}

We first define the communication-sensing (C-S) channel correlation and reveal the relationship between this metric and the dual-functional performance.
The C-S channel correlation is measured by the distance between the power distributions in the beamspace. The communication and sensing channels are first transformed into the beamspace through Discrete Fourier Transform (DFT) as
\vspace{-0.1cm}
\begin{equation}
    \bm h^{b}_{\text{c},m}=\sum_{u=1}^{U}\bm D_{t}^{\text{H}}\bm h_{u,m},~~\bm h^{b}_{\text{s},m}=\text{diag}(\bm D_{r}^{\text{H}}\bm G_{m}\bm D_{t}),
    \vspace{-0.1cm}
\end{equation}
where $\bm D_{t}\in \mathbb{C}^{N_{t}\times N_{D}}$ and $\bm D_{r}\in \mathbb{C}^{N_{r}\times N_{D}}$ are the dictionary uniformly sampled in beamspace \cite{shijian_MI_max}, with the number of codewords being $N_D$. For communications, the beamspace channels for all users are aggregated.
It is also worth mentioning that $\bm G_{m}$ consists of the steering vectors at both ends. Consequently, the transmitter and receiver are separately transformed into the beamspace. The beamspace channels across all subcarriers are then aggregated and normalized as
\vspace{-0.1cm}
\begin{equation}
    \widehat{\bm h}^{b}_{\text{c}}= \frac{\sum_{m=1}^{M}\vert \bm h^{b}_{\text{c},m} \vert}{\| \sum_{m=1}^{M}\vert \bm h^{b}_{\text{c},m} \vert \|_{1}},~~\widehat{\bm h}^{b}_{\text{s}}=\frac{\sum_{m=1}^{M}\vert \bm h^{b}_{\text{s},m} \vert}{\| \sum_{m=1}^{M}\vert \bm h^{b}_{\text{s},m} \vert \|_{1}}.
\end{equation}
Note that they can also be interpreted as the discrete probability distribution for communication and sensing in the beamspace.
With $\widehat{\bm h}^{b}_{\text{c}}$ and $\widehat{\bm h}^{b}_{\text{s}}$ defined, the C-S channel correlation is quantified as follows:
\vspace{-0.15cm}
\begin{equation}
    \text{Cor}(\bm H, \bm G)=\frac{1}{\text{KL}( \widehat{\bm h}^{b}_{\text{c}}, \widehat{\bm h}^{b}_{\text{s}}  )},
    \vspace{-0.1cm}
\end{equation}
where $\text{KL}(\bm p_1, \bm p_2)=\sum_{n=1}^{N} p_1[n]\ln \frac{p_1[n]}{p_2[n]}$ denotes the Kullback-Leibler (K-L) divergence between distributions $\bm p_1$ and $\bm p_2$, $\bm H=\{\bm h_{u,m}\}_{u=1,m=1}^{U,M}$ and $\bm G=\{\bm G_{m}\}_{m=1}^{M}$ denote the communication and sensing channels for all users across all subcarriers.

\vspace{-0.15cm}
\begin{proposition} 
As the C-S channel correlation $\text{Cor}(\bm H, \bm G)$ increases, the SE improves, and the CRB decreases in a monotonic way.
\end{proposition}
\vspace{-0.2cm}
\begin{proof}
See Appendix A.
\end{proof}
\setlength{\parindent}{0pt}
\textbf{Remark 1.} 
The C-S channel correlation is an intrinsic property of ISAC transmission. From Proposition 1, establishing equivalent channels with higher correlation facilitates the rotation of the communication and sensing subspaces, as shown in \cite{RIS_rotation_ISAC}, potentially expanding the Pareto performance boundary for ISAC systems. Without the need for additional external components, the modulation of the equivalent channel can be achieved using the TTDs connected to the UPA, i.e., $\widetilde{\bm h}_{\text{c},m}=\bm F_{\text{TD,m}}^{\text{H}}\bm h_{\text{c},m}$ and $\widetilde{\bm G}_{m}=\bm G_{m}\bm F_{\text{TD,m}}$.
Thus, the role of the TTDs is to maximize the C-S channel correlation, which will guide the hybrid precoder design in the next section.

\setlength{\parindent}{10pt}

\section{SoM-enhanced prior acquisition}

The current ISAC framework relies on arduous efforts on obtaining prior CSI and optimizing the precoder. However, the substantial antenna arrays and dense subcarrier configurations inherent to sub-THz systems impose considerable latency during CSI acquisition and precoder design phases. To address these challenges, this section proposes a sensor-aided strategy that synergizes out-of-band visual data with sparse RF measurements to augment operational efficiency.

\vspace{-0.1cm}
\subsection{Frame Structure}

\begin{figure*}[t]
  \vspace{-0.0cm}
  \centering
  \includegraphics[width=0.88\linewidth]{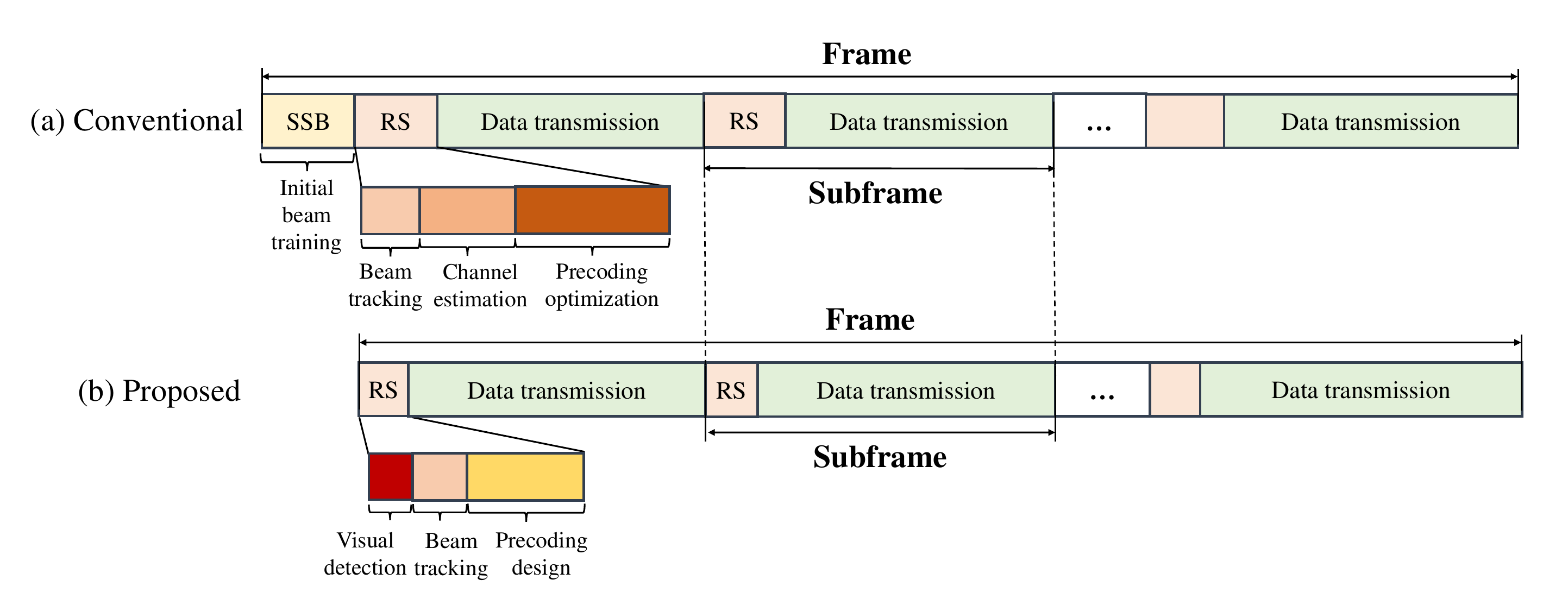}
  \vspace{-0.2cm}
  \captionsetup{font=small}
  \caption{Frame structure comparison between RF-only and vision-aided schemes.}
  \label{fig:frame structure}
  \vspace{-0.5cm}
\end{figure*}

According to 3GPP Release 18 \cite{5GNR_frame, 3GPP_R18}, each new radio (NR) frame is divided into $N_{\text{sub}}$ subframes. The positions of mobile users and targets are assumed to be invariant within one subframe. As illustrated in Fig.~\ref{fig:frame structure}, each subframe consists of a reference signal (RS) stage and a data transmission stage. 
\begin{itemize}
    \item Reference signal: In the RS stage, visual information is first used to detect candidate users and targets, followed by a fast beam tracking process to refine communication user angles. These estimations are then used to generate a positioning spectrum containing channel-related parameters. The positioning spectrum and images are input into a pre-trained neural network to predict the hybrid precoding parameters for ISAC.
    \item Data transmission: Once the precoding is designed, the communication symbols are precoded and transmitted to the users during the remaining time of the subframe, while the targets are estimated through echo reception.
\end{itemize}

As demonstrated in Fig.~\ref{fig:frame structure}, existing frame structures in RF-only transmission protocols, such as \cite{beam_tracking_tanjingbo} and \cite{dynamic_MP}, require an additional synchronization signal block (SSB) at the beginning of the frame for initial beam training. In contrast, the proposed ISAC transmission frame eliminates the need for both initial beam training and instantaneous channel estimation, resulting in simpler operations in the RS.

\vspace{-0.2cm}
\subsection{Identification of Candidate User and Target}
Unlike beam training via RF measurements only, we leverage the visual information to instantaneously estimate potential users' angular ranges.
To accommodate the rapidly changing environment, an efficient visual processing method is essential. To this end, we employ YOLO-v5-s, a widely used lightweight object detector in the computer vision field.

During the offline stage, we pre-train a customized YOLO-v5-s model on our dataset, by minimizing the MSE between the network's output and the annotation labels, enabling it to accurately detect the candidates and identify them as users or targets.
During the deployment stage, the RGB image captured is input into this pre-trained detector, so the corresponding bounding boxes of the users and targets are obtained as
\vspace{-0.1cm}
\begin{equation}
    \bm b_{i}=[b_{x,i}, b_{y,i}, b_{w,i}, b_{h,i}, c_{i}],
    \vspace{-0.1cm}
\end{equation}
where $[b_{x,i}, b_{y,i}]$ (in pixels) signifies the center of the $i$-th bounding box. $b_{w,i}$ and $b_{h,i}$ (in pixels) denote its width and height, and $c_{i}$ denotes the category (0 for the user and 1 for the target).
Subsequently, based on the principle of pinhole imaging \cite{pinhole}, we transform the points within the candidate bounding boxes into the world coordinates as
\vspace{-0.1cm}
\begin{equation}
    \bm p_{\text{wd},i}=\bm R_{j}^{-1}( z_{\text{cam},i} \bm K^{-1} \bm p_{\text{img},i} - \bm p_{\text{cam}}),
    \label{equ:calibration}
    \vspace{-0.1cm}
\end{equation}
where $\bm p_{\text{wd},i}=[x_{\text{wd},i}, y_{\text{wd},i}, z_{\text{wd},i}]^{\text{T}}$ denotes the world coordinates of the corresponding center point $\bm p_{\text{img},i}=[b_{x,i}, b_{y,i}, 1]^{\text{T}}$. $z_{\text{cam},i}$ is the depth of this point according to the depth map. $\bm p_{\text{cam}}$ is the world coordinates of the camera. 
In the above transformation process,
\vspace{-0.1cm}
\begin{equation}
\bm K=
\begin{bmatrix}
&\frac{N_{w}}{2\tan(fov/2)} , &0, &\frac{N_{w}}{2} \\
& 0, &\frac{N_{w}}{2\tan(fov/2)}, &\frac{N_{h}}{2} \\
& 0 , &0, & 1 
\end{bmatrix},
\vspace{-0.1cm}
\end{equation}
represents the intrinsic matrix of the camera determined by its configuration. $\bm R= \bm R_{z}\bm R_{y}\bm R_{x}$ denotes the rotation matrix, with
\vspace{-0.1cm}
\begin{equation}
\bm R_{x}
=
\begin{bmatrix}
&1 , &0, &0 \\
& 0, &\cos \gamma_{x}, &-\sin\gamma_{x}  \\
& 0 , &\sin\gamma_{x}, &\cos\gamma_{x} 
\end{bmatrix},
\tag{17a}
\end{equation}
\begin{equation}
\bm R_{y}
=
\begin{bmatrix}
&\cos\gamma_{y} , &0, &\sin\gamma_{y} \\
& 0, &1, &0 \\
&-\sin\gamma_{y} , &0, &\cos\gamma_{y} 
\end{bmatrix},
\tag{17b}
\end{equation}
\begin{equation}
\bm R_{z}
=
\begin{bmatrix}
&\cos\gamma_{z} , &-\sin\gamma_{z}, &0 \\
&\sin\gamma_{z}, &\cos\gamma_{z}, &0 \\
&0 , &0, &1 
\end{bmatrix},
\tag{17c}
\end{equation}
where $\gamma_{x}$, $\gamma_{y}$ and $\gamma_{z}$ denote the angles of the camera with respect to the x, y and z-axis. Denote $\bm p_{\text{UPA}}=[x_{\text{UPA}}, y_{\text{UPA}}, z_{\text{UPA}}]^{\text{T}}$ as the world coordinates of the UPA.
Then we can obtain the candidate's polar coordinates with respect to the UPA, $[\widehat{\phi}_{i}, \widehat{\theta}_{i}, \widehat{d}_{i}]$, as
\vspace{-0.1cm}
\begin{align}
    &\widehat{\phi}_{i}\!=\!\arctan \left( \frac{y_{\text{wd},i}-y_{\text{UPA}}}{x_{\text{wd},i}-x_{\text{UPA}}} \right), \tag{18a} \\
    &\widehat{\theta}_{i,j}\!=\!\pi\!\!-\!\arctan\!\!\left( \!\!\frac{\sqrt{\!(x_{\text{wd},i}\!-\!x_{\text{UPA}})^{2}\!+\!(y_{\text{wd},i}\!-\!y_{\text{UPA}})^{2}}}{\vert z_{\text{wd},i}-z_{\text{UPA}}\vert } \right)\!, \tag{18b} \\
    &\widehat{d}_{i}\!=\!\| \bm p_{\text{wd},i} - \bm p_{\text{UPA}} \|_{2}. \tag{18c}
    \vspace{-0.1cm}
\end{align}
The angular ranges of the candidates relative to the UPA can be determined by calculating the polar coordinates of the bounding box vertices in a similar manner. Therefore, the horizontal and vertical angular ranges of the users along with their distances can be obtained as $\{[\hat{\phi}_{\text{c},u,\text{min}},\hat{\phi}_{\text{c},u,\text{max}},\hat{\theta}_{\text{c},u,\text{min}},\hat{\theta}_{\text{c},u,\text{max}},\hat{d}_{\text{c},u}]_{u=1}^{U}\}$
\footnote{
Note that in real-world scenarios, multiple cameras may be deployed to enhance coverage. The proposed detection scheme is also applicable to multi-camera setups. Specifically, the detection results from each camera can be fused based on its range in world coordinates and its category, thereby improving the accuracy of overlapping regions \cite{multi_view_1}. Additionally, the portion of the camera's view outside the UPA coverage will not be considered.
}.

Then the angular ranges of targets can be determined in a similar manner, which can be expressed as $\mathcal{T}=\{[\hat{\phi}_{\text{s},k,\text{min}},\hat{\phi}_{\text{s},k,\text{max}},\hat{\theta}_{\text{s},k,\text{min}},\hat{\theta}_{\text{s},k,\text{max}},\hat{d}_{\text{s},k}]_{k=1}^{K}\}$.

\vspace{-0.2cm}
\subsection{Squint-Aware Fine Beam Tracking}

\begin{figure*}[t]
\vspace{-0.0cm}
     \begin{align}
     \!\!\bm T[q_{{h}},\!q_{{v}}]\!\!&=\!\!\frac{1}{L_{h}L_{v}}\!\!\sum_{n_{t_{h}}\!\!\!=\!(\!q_{h}\!-\!1\!)\!L_{h}\!\!+\!1}^{q_{h}L_{h}}\sum_{n_{t_{v}}\!\!=\!(\!q_{v}\!-\!1\!)\!L_{v}\!\!+\!1}^{q_{h}L_{h}}\!\!
    \frac{f_1}{\!2Bf_\text{c}\!}\!\biggl( \!\!\sin\theta_{0}\!\sin\!\phi_{0}(\!n_{t_{h}}\!\!\!-\!1\!)\!+\! \cos\!\theta_{0}(\!n_{t_{v}}\!\!\!-\!1\!)\!\!\biggr)
    \!\!-\!\!\frac{f_M}{\!2Bf_{\text{c}}\!}\!\biggl(\!\! \sin\!\theta_{1}\!\sin\!\phi_{1}(\!n_{t_{h}}\!\!\!-\!1\!) 
    \!+\! \cos\!\theta_{1}(\!n_{t_{v}}\!\!\!-\!1\!)\!\! \biggr)\!, \tag{19} \label{equ:CBS_TTD}  \\
    \!\bm \varphi[n_{t_{h}}\!,\!n_{t_{v}}]\!\!&=2f_{1}T[q_{{\text{h}}},q_{{\text{v}}}]+\frac{f_1}{f_c} \biggl( \sin\theta_{0}\sin\phi_{0}(n_{t_{h}}-1)   
    + \cos\theta_{0}(n_{t_{v}}-1) \biggr). \tag{20} \label{equ:CBS_PS}
    \vspace{-0.1cm}
\end{align}
\vspace{-0.05cm}
\hrulefill
\vspace{-0.35cm}
\end{figure*}

In sub-THz systems, visual information alone is insufficient for accurately pointing extremely narrow beams. To improve the accuracy of the prior information, incorporating a small number of RF operations is necessary.

Based on the angular ranges obtained from visual detection, we propose a squint-aware cross-pattern beam tracking (SA-CP-BT) scheme for refining angle estimation. Current techniques like \cite{beam_tracking_tanjingbo, beam_tracking_shixu}, require a complete scan. In contrast, SA-CP-BT only needs to cover the cross-pattern tracks within each rectangle, thereby reducing time overhead.

Specifically, the beams across all subcarriers can be zoomed in from $(\phi_{0}, \theta_{0})$ to $(\phi_{1}, \theta_{1})$ by setting TTDs and PSs as Eq. (\ref{equ:CBS_TTD}) and Eq. (\ref{equ:CBS_PS}). The pointing angles on each subcarrier can be arranged in sequence along this trajectory.
Building on this, we can effectively control the beam squint to cover the desired angular trajectory. Nevertheless, solely relying on beam squint to cover all angles across all ranges remains time-consuming.

Consider a user located at $(\phi_{u},\theta_{u})$ within an angular range, where the width is $\Delta\phi$ and the height is $\Delta \theta$.
The beam squint is controlled to traverse this range along a specific trajectory, pointing to $(\phi_{m},\theta_{m})$ at the $m$-th subcarrier. The angular distance between the user and $(\phi_{m},\theta_{m})$ can be computed as $d(\phi_{u},\theta_{u},m)=\left ((\phi_{u}-\phi_{m})^{2}+(\theta_{u}-\theta_{m})^{2}\right )$.

\begin{proposition}
    If the angular range satisfies $\Delta\theta\leqslant \Delta^{\text{SA}}_{\theta}=\frac{4f_{\text{c}}}{N_{t_{h}}f_{M}}$ and $\Delta \phi\leqslant \Delta^{\text{SA}}_{\phi}=\frac{4f_{\text{c}}}{N_{t_{v}}f_{M}}$, the array gain at different subcarriers along this beam squint trajectory decreases with $d(\phi_{u},\theta_{u},m)$.
\end{proposition}
\vspace{-0.15cm}
\begin{proof}
    See Appendix B.
\end{proof}
\begin{lemma}
When squint beams horizontally traverse an angular range at a height of $\Delta\theta\leqslant \Delta^{\text{SA}}_{\theta}$,  
the user azimuth angle in this range can be unambiguously resolved by identifying the subcarrier exhibiting maximal array gain.
\end{lemma}

Drawing on Proposition 2 and Lemma 1, within a sufficiently small angular range, when squint beams simply traverse this range along a straight line in a given direction, there is no need to scan other angles within this region. Instead, the actual angle of the user in this direction can be accurately determined based on the current beam responses. In contrast, if this range is exceeded, angle ambiguity may occur, meaning that the angle corresponding to the subcarrier with the largest response may not represent the user's true angle.

Based on the above analysis, we can simply control the beam squint in a cross-pattern manner to refine the angle estimate. The SA-CP-BT scheme first partitions the coarse-grained angular range to satisfy the condition in Lemma 1, and then performs horizontal and vertical searches to sequentially identify the azimuth and elevation angles.

\begin{figure}[t]
  \vspace{-0.0cm}
  \centering
  \includegraphics[width=1.0\linewidth]{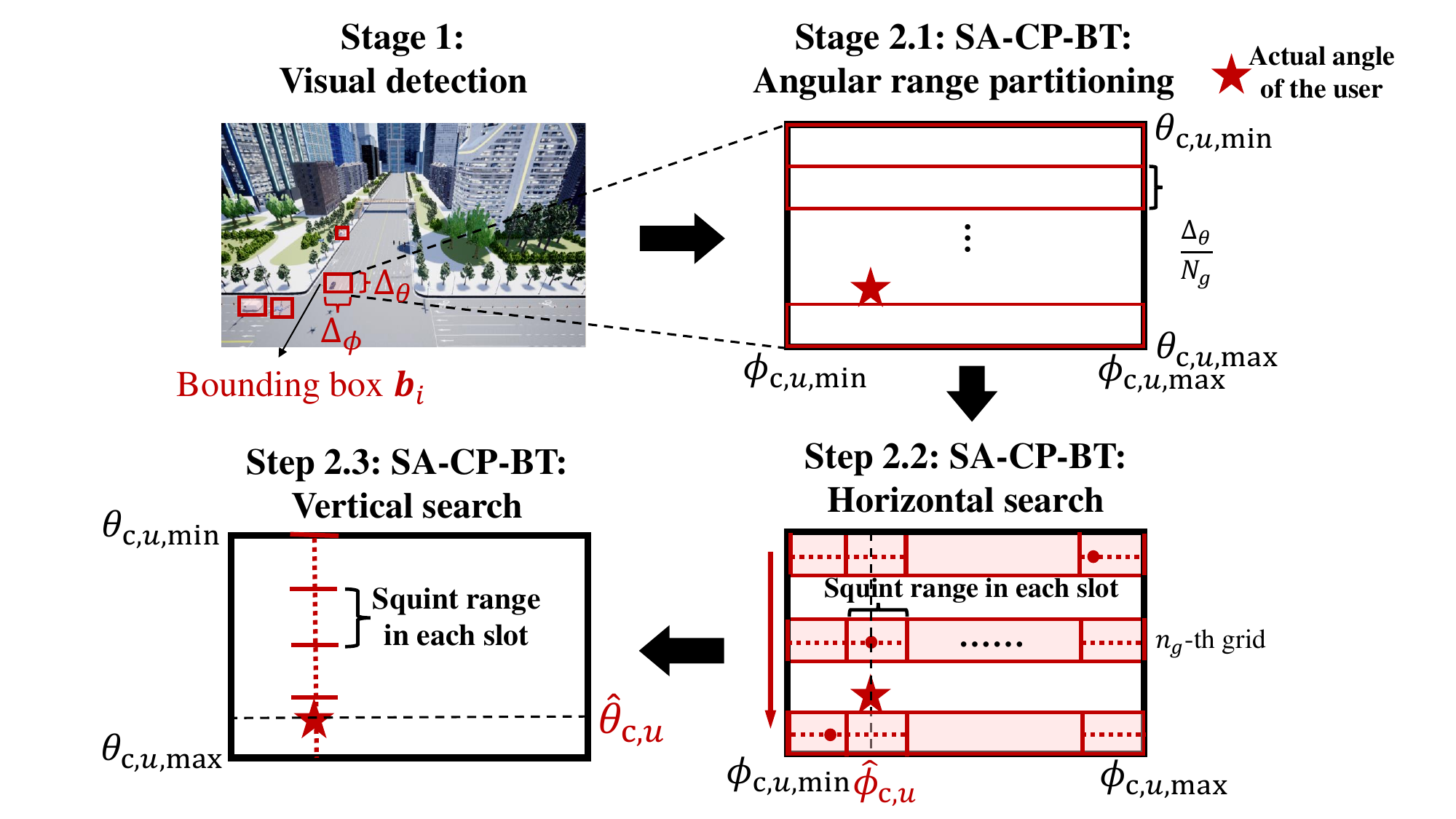}
  \vspace{-0.2cm}
  \captionsetup{font=small}
  \caption{An illustration of the initial beam training and the proposed beam tracking scheme.}
  \label{fig:beam training}
  \vspace{-0.4cm}
\end{figure}

\subsubsection{Angular Range Partitioning}

We first divide the angular range vertically into $N_{g}$ grids, such that $\frac{\hat{\theta}_{\text{c},u,\text{max}}-\hat{\theta}_{\text{c},u,\text{min}}}{N_{g}}\leqslant \Delta_{\theta}^{\text{SA}}$.

\subsubsection{Horizontal search}

In each angular grid, the beam squint is first controlled for the horizontal search. 
Denote $T_{bt}$ as the number of time slots used for tracking in each angular grid. In each slot, the beams cover the horizontal trajectory from $(\phi_{0},\theta_{0})$ to $(\phi_{1},\theta_{1})$, with $\phi_{0}=\hat{\phi}_{\text{c},u,\text{min}}+((t_{bt}-1)\Delta_{\phi}/T_{bt})$, $\phi_{1}=\hat{\phi}_{\text{c},u,\text{min}}+(t_{bt}\Delta_{\phi}/T_{bt})$, and $\theta_{0}=\theta_{1}=(\hat{\theta}_{\text{c},u,\text{min}}+\hat{\theta}_{\text{c},u,\text{max}})/2$. To this end, the TTDs and the PSs are set as (\ref{equ:CBS_TTD}) and (\ref{equ:CBS_PS}) respectively. The transmitted probing signal is simply set to $s_{m}=1$.
Then the received signal at the $u$-th user at the $m$-th subcarriers in the $t_{bt}$-th slot in the $n_g$-th grid is
\vspace{-0.1cm}
\begin{equation}
     y^{(h,u,t_{bt},n_{g})}_{m}=\bm h_{u,m}^{\text{H}}\bm F^{(h,u,t_{bt},n_{g})}_{\text{TD},m}\bm f^{(h,u,t_{bt},n_{g})}_{\text{PS}}\!+\! z_{\text{c},u,m}.
     \vspace{-0.1cm}
     \tag{21}
\end{equation}
The user feeds back to the BS the subcarrier and slot index at which the modulus of the received signal reaches its maximum across all slots, i.e., $(m_{h}^{*},t_{bt,h}^{*},n_{g}^{*})=\text{argmax}_{m,t_{bt},n_{g}} \vert y^{(h,u,t_{bt},n_{g})}_{m}\vert$.
Then the azimuth of the $u$-th user can be calculated as
\vspace{-0.1cm}
\begin{equation}
\begin{split}
\hat{\phi}_{\text{c},u} = &\arcsin \biggl[ \frac{(f_{M}\! -\! f_{m_{h}^{*}})f_{1}}{Bf_{m_{h}^{*}}} \sin \left( \hat{\phi}_{\text{c},u,\text{min}}\! +\! ((t_{bt}^{*}\!-\!1)\Delta_{\phi}/T_{bt}) \right) \\
&~- \frac{f_M(f_{1} - f_{m_{h}^{*}})}{f_{m_{h}^{*}}B} \sin \left( \hat{\phi}_{\text{c},u,\text{min}} + (t_{bt}^{*}\Delta_{\phi}/T_{bt}) \right) \biggr].
\end{split}
\raisetag{4ex} 
\label{eq:example}
\tag{22}
\end{equation}

\subsubsection{Vertical search}
With the estimated azimuth angle, we initiate squint-aware tracking along the vertical direction during $T_{bt}$ slots, by setting $\phi_{0}=\phi_{1}=\hat{\phi}_{c,u}$, $\theta_{0}=\hat{\theta}_{\text{c},u,\text{min}}+((t_{bt}-1)\Delta_{\theta}/T_{bt})$ and $\theta_{1}=\hat{\theta}_{\text{c},u,\text{min}}+(t_{bt}\Delta_{\theta}/T_{bt})$ in the $t_{bt}$-th slot. The user then returns the subcarrier and slot index as $(m_{v}^{*}, t_{bt,v}^{*})=\text{argmax}_{m,t_{bt}} \vert y^{(v,u,t_{bt})}_{m}\vert$.
Eventually, the elevation angle of the $u$-th user can be determined as
\begin{equation}
\begin{split}
\hat{\theta}_{\text{c},u} &=\! \arccos \biggl[ \frac{(f_{M}\!-\!f_{m_{\text{v}}^{*}})f_{1}}{Bf_{m_{\text{v}}^{*}}} \cos \left(\hat{\theta}_{\text{c},u,\text{min}} \!+ \!((t_{bt}^{*}\!-\!1)\Delta_{\theta}/T_{bt}) \right) \\
&~~~~~~~~~\quad -\frac{f_M}{f_{m_{\text{v}}^{*}}} \cos \left(\hat{\theta}_{\text{c},u,\text{min}} + (t_{bt}^{*}\Delta_{\theta}/T_{bt}) \right) \biggr].
\end{split}
\raisetag{4ex} 
\label{eq:example_1}
\tag{23}
\end{equation}
Through SA-CP-BT, the accurate angles of the users, represented as $\mathcal{U}=\{(\hat{\phi}_{\text{c},u},\hat{\theta}_{\text{c},u},\hat{d}_{\text{c},u})_{u=1}^{U}\}$, can be obtained.

\section{SoM-enhanced ISAC Precoding}

\begin{figure*}[t]
  \vspace{0.0cm}
  \centering
  \includegraphics[width=0.90\linewidth]{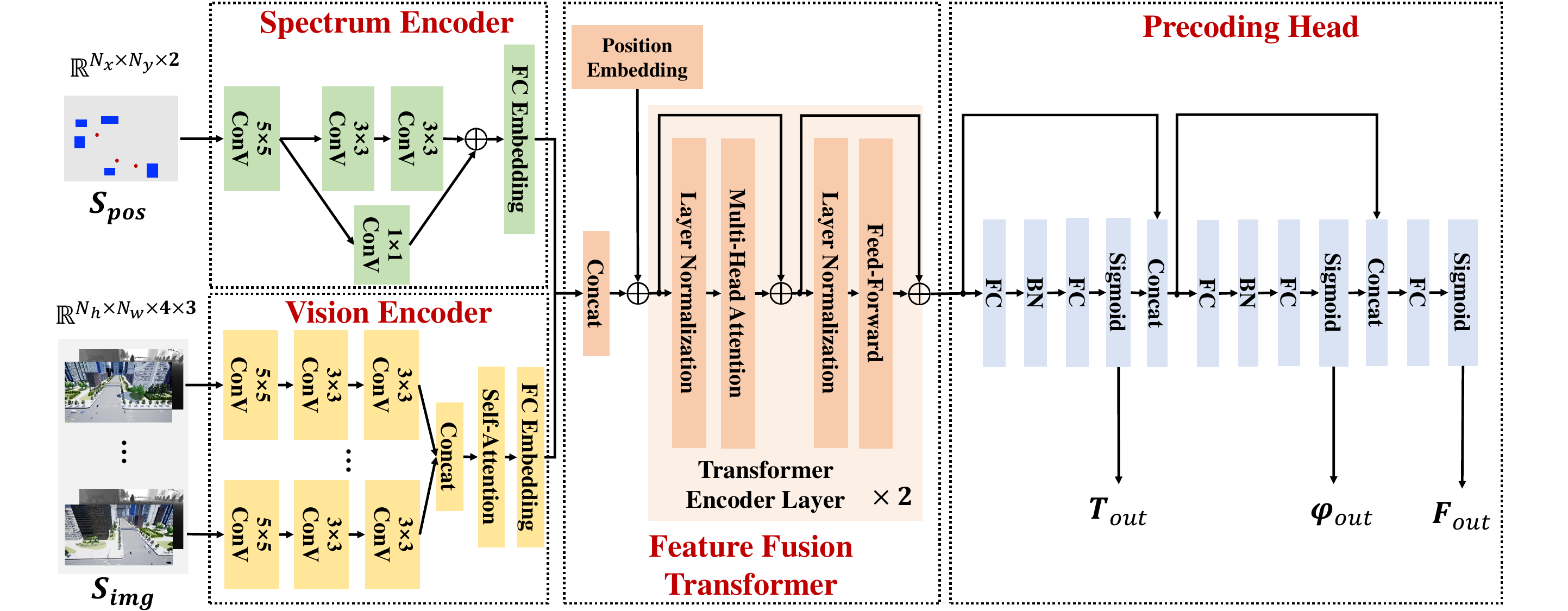}
  \vspace{-0.05cm}
  \captionsetup{font=small}
  \caption{An illustration of the architecture of the proposed ViR-Net.}
  \label{fig:network structure}
  \vspace{-0.25cm}
\end{figure*}

Based on the prior information obtained from the aforementioned steps, we devise an unsupervised learning-assisted approach to jointly design the parameters of ISAC hybrid precoding. In this approach, both the RF-based prior information and visual information will be utilized to enhance the dual-functional performance.

\subsection{Data pre-processing}

The input data contains both the users' and targets' positioning prior information represented by $\bm S_{pos}\in \mathbb{R}^{N_{x}\times N_{y}\times 2}$ and the RGB-D images captured from cameras represented by $\bm S_{img}\in \mathbb{R}^{N_{h}\times N_{w}\times 4\times 3}$.
\vspace{-0.00cm}
\begin{itemize}
    \item RF-related positioning prior: Once the angular prior $[\hat{\phi}_{c,u},\hat{\theta}_{c,u}]_{u=1}^{U}$, $[\hat{\phi}_{\text{s},k,\text{min}},\hat{\phi}_{\text{s},k,\text{max}},\hat{\theta}_{\text{s},k,\text{min}},\hat{\theta}_{\text{s},k,\text{max}}]_{k=1}^{K}$ and depth prior $\{\widehat{d}_{\text{c},u}\}_{u=1}^{U}, \{\widehat{d}_{\text{s},k}\}_{k=1}^{K}$ are obtained through visual detection and fine beam tracking, they are transformed into a 3D positioning spectrum $\bm S_{pos}$ as
    \vspace{-0.2cm}
    \begin{equation}
        \!\bm S_{pos}[n_{x}, n_{y},1]\! = \!
  \begin{cases}
    1\!-\!\frac{\hat{d}_{\text{c},u}}{d_{\text{max}}}\!, &\text{if $\mathcal{S}_{n_{x},n_{y}} \bigcap \mathcal{U} \!\neq \!\emptyset$},\\
    0,\!\! &\text{Otherwise},
  \end{cases}
  \nonumber
    \end{equation}
    \vspace{-0.15cm}
    \begin{equation}
        \!\bm S_{pos}[n_{\text{x}}, n_{\text{y}},2] \!= \!
  \begin{cases}
    1\!-\!\frac{\hat{d}_{\text{s},k}}{d_{\text{max}}}, &\text{if $\mathcal{S}_{n_{x},n_{y}} \bigcap \mathcal{T} \!\neq \!\emptyset$},\\
    0,\!\! &\text{Otherwise}.
  \end{cases}
  \nonumber
  \vspace{-0.1cm}
    \end{equation}
where $\mathcal{S}_{n_{x},n_{y}}=\{[\theta_{n_{x}},\!\theta_{n_{x}\!+\!1})\!\!\times \!\! [\phi_{n_{y}},\! \phi_{n_{y}\!+\!1})\}$ denotes the quantized angular grid, with $\theta_{n_{x}}\!=\frac{\pi}{N_{x}}(n_{x}-1)$, $\phi_{n_{y}}\!=\!-\frac{\pi}{2}\!+\!\frac{\pi}{N_{y}}(n_{y}-1)$. $d_{\text{max}}$ defines the maximal detectable distance within the BS's coverage.
    \item Visual information: At each camera, the RGB image and the depth map are concatenated to form $\bm I_{vi}^{(j)}=[\bm I^{(j)}_{rgb}, \bm I^{(j)}_{dep}]\in \mathbb{R}^{N_{h}\times N_{w} \times 4}$ ($j\!\in\!\{l,m,r\}$), followed by the normalization operation. Consequently, the RGB-D images from three perspectives, i.e., $\bm S_{img}=[\bm I_{vi}^{(l)}, \bm I_{vi}^{(m)},\bm I_{vi}^{(r)}]$, are fed into the network.
\end{itemize}

\subsection{Network Architecture}

We devise a vision-RF fusion network for squint-aware hybrid precoder design, which is termed as ViR-Net. 
The network model consists of four components, including a spectrum encoder (SpE), a vision encoder (ViE), a feature-fusion transformer (FFTF), and a precoding head, as demonstrated in Fig.~\ref{fig:network structure}.

\subsubsection{\textbf{Spectrum Encoder}}

The channel-related features from the positioning spectrum $\bm S_{pos}$ are extracted by SpE. One layer of convolutional blocks is utilized, comprising one $5\times 5$ convolutional layer, one Gaussian error linear unit (GELU) activation function, and one batch normalization (BN) layer. Subsequently, a residual block is appended, which incorporates two $3\times 3$ convolutional layers along with a $1\times 1$ skip convolutional layer. Then a fully connected (FC) embedding layer projects the features into a unified feature space, yielding $\bm V_{pos}\in \mathbb{R}^{L_{f}\times d_{pos}}$, where $L_{f}$ is the feature space dimension and $d_{pos}$ is the number of features.

\subsubsection{\textbf{Vision Encoder}}

Since $\bm S_{pos}$ only encompasses the positioning prior for the LoS paths for communications, the information regarding the non-LoS paths is overlooked. Therefore, we exploit RGB-D images to implicitly capture the prior of non-LoS paths.

ViE processes RGB-D images to extract channel-related features as a supplement to SpE. For each perspective, ViE consists of three convolutional blocks, each with a convolutional layer, GELU, max-pooling, and BN. Then features from the three perspectives are concatenated, followed by a self-attention mechanism for visual feature fusion. An FC embedding layer outputs $\bm V_{img}\in \mathbb{R}^{L_{f}\times d_{img}}$.
To expedite training, we use a self-supervised strategy to pre-train the convolutional blocks of ViE.  Specifically, we build an auto-encoder to compress and recover RGB-D images, with the encoder formed by the ViE convolutional blocks, as detailed in \cite{selfsuper_1}. After training, the encoder is isolated to construct ViE.

\subsubsection{\textbf{Feature Fusion Transformer}}

We utilize the Transformer structure to globally extract the correlation between the features from the positioning spectrum and the images.
$\bm V_{pos}$ and $\bm V_{img}$ are first concatenated into a sequence of feature vectors with dimension $L_{f}\times (d_{pos}+d_{img})$. Then the position embedding is used to represent the relative positions \cite{PE}. Subsequently two layers of Transformer encoder, each consisting of a multi-head self-attention sub-layer and a multi-layer perceptron sub-layer.

\subsubsection{\textbf{Precoding Head}}

The precoding head projects the fused features into the values of TTDs, PSs and the digital precoder sequentially.
Two FC layers are first employed to predict the normalized value of TTDs as $\hat{\bm t}\in [0,1]^{Q_{t}}$, with the Sigmoid activation function. The output is $\bm T_{\text{out}}=\text{reshape}(t_{\text{max}}\hat{\bm t})$.
Concatenating fused features and $\hat{\bm t}$ as the input, we proceed to predict PSs through two FC layers, and the output is $\hat{\bm \varphi}\in [0,1]^{N_{t}\times N_{\text{RF}}}$. Then PSs are calculated as $\bm \varphi_{\text{out}}=2\pi\hat{\bm \varphi}$. 
Finally, we concatenate fused features, $\hat{\bm t}$ and $\hat{\bm \varphi}$ to predict the digital part using another FC layer. The output $\hat{\bm f}\in \mathbb{R}^{2N_{\text{RF}}^{2}M\times 1}$ is reshaped as the complex matrices $\bm F_{\text{out}}\in \mathbb{C}^{N_{\text{RF}}\times N_{\text{RF}}\times M}$, and then normalized to satisfy the power constraint.

\begin{table}[t]
\vspace{-0.0cm}
    \centering
    \caption{System setup for the transceiver's hardware}
    \begin{tabular}{p{1.6cm}p{1.6cm}||p{1.6cm} p{1.6cm}}
    \hline
       Parameter  &  value & Parameter  &  value \\
       \hline
       \hline
       $N_{t}$  & 256   &  $N_{r}$  & 256   \\
       \hline
       $N_{t_{h}}$ & 16   &  $N_{t_{v}}$  & 16   \\
       \hline
       $Q_{t_{h}}$ &  16  & $Q_{t_{v}}$  & 16   \\
       \hline
       $N_{\text{RF}}$ &  6  &  $N_{D}$ &  4096  \\
       \hline
       $t_{\text{max}}$ &  1 ns  &  $T_{bt}$ &  1  \\
       \hline
       Pilots &  16  & $N_{\text{sub}}$  & 10    \\
       \hline
    \end{tabular}
    \label{tab:hardware}
\end{table}
\vspace{-0.1cm}

\begin{table}[t]
    \centering
    \caption{System setup for the sub-THz channels}
    \begin{tabular}{p{1.6cm}p{1.6cm}||p{1.6cm} p{1.6cm}}
    \hline
       Parameter  &  value & Parameter  &  value \\
       \hline
       \hline
       $f_{\text{c}}$  & 100 GHz   &  $B$  & 8 GHz   \\
       \hline
       $M$ & 32   &  $P_{u}$  & 4   \\
       \hline
       $K_{f}$ &  4  & $\sigma_{\text{RCS}}$  & 1   \\
       \hline
       $U$ &  6 GHz  &  $K$ &  3  \\
       \hline
       $n_{\text{c},0}$ &  -30 dBm/Hz  &  $n_{\text{s},0}$ &  -30 dBm/Hz  \\
       \hline
    \end{tabular}
    \label{tab:channel}
\end{table}
\vspace{-0.2cm}

\begin{table}[t]
\vspace{-0.0cm}
    \centering
    \caption{System setup for the physical environment}
    \begin{tabular}{p{1.8cm}p{1.75cm}||p{1.8cm} p{1.75cm}}
    \hline
       Parameter  &  value & Parameter  &  value \\
       \hline
       \hline
       UPA's height  & 55 m   &  UAVs' height  &  45 m   \\
       \hline
       Users' height & 1.4 m   &  $fov$  & 100 deg   \\
       \hline
       $N_{w}$ &  1920  & $N_{h}$  & 1024   \\
       \hline
       Cam. 1 rel. pos. &  (5, 0, 0.1) m  &  Cam. 1 rot. ang. &  (0, 25, -90) deg  \\
       \hline
       Cam. 2 rel. pos. &  (4.8, 0, 0.1) m  &  Cam. 2 rot. ang. &  (0, 25, 0) deg  \\
       \hline
       Cam. 3 rel. pos. &  (4.9, 0, 0.1) m  &  Cam. 3 rot. ang. &  (0, 25, 90) deg  \\
       \hline
    \end{tabular}
    \label{tab:visual}
\end{table}
\vspace{-0.1cm}

\begin{table}[t]
    \centering
    \caption{Hyper-parameters for model training}
    \begin{tabular}{p{1.8cm}p{1.6cm}||p{1.8cm} p{1.6cm}}
    \hline
       Parameter  &  value & Parameter  &  value \\
       \hline
       \hline
       Optimizer  & Adam   &  Learning rate   & 0.004   \\
       \hline
       $N_{\text{b}}$ & 16   &  $L_{f}$  & 256   \\
       \hline
       $d_{pos}$ &  256  & $d_{img}$  & 256   \\
       \hline
    \end{tabular}
    \label{tab:model}
\end{table}
\vspace{-0.2cm}

\vspace{-0.1cm}
\subsection{Loss function}

We employ an unsupervised learning-based method for the model training. To achieve the goal of maximizing ISAC performance with the awareness of the C-S channel correlation, we customize the loss function for one batch as
\vspace{-0.1cm}
\begin{equation}
\begin{aligned}
    L\!&=-\frac{1}{N_{\text{b}}}\!\sum_{i=1}^{N_{\text{b}}}\frac{\text{Cor}(\bm H^{(i)}, \bm G^{(i)})}{\text{Cor}^{\star}(\bm H^{(i)}, \bm G^{(i)})}\times\\
    & ~~~~\left[
    \frac{\mathcal{CRB}_{\text{min}}^{(i)}}{\mathcal{CRB}^{(i)}}-\eta_{\text{c}} \text{ReLU}(\Gamma\!-\!\mathcal{R}^{(i)})\!\right ],
    \label{equ:loss}
    \vspace{-0.1cm}
    \end{aligned}
   \tag{24}
\end{equation}
where $\phi_{\text{ViR}}$ denote the parameters of ViR-Net, $N_{\text{b}}$ is the batch size, $\Gamma$ denotes the SE threshold for communications and $\eta_{\text{c}}$ is the weighting coefficient.
In this loss function, the objective of model training is twofold: To enhance the C-S channel correlation for modulating the communication and sensing subspaces, and to separately improve the performance of communication and sensing, thereby expanding the performance boundaries of ISAC.
In the offline training stage, ViR-Net is updated with this loss function by backpropagation in randomly sampled batches. In the deployment, the positioning spectrum and the RGB-D images are captured and input to the pre-trained ViR-Net, and then the values of TTDs, PSs and digital precoding are obtained sequentially.

\section{Simulations}

In this section, experimental results are presented to evaluate the performance of the proposed schemes. 

\subsection{Scenario Construction}

We construct a low-altitude ISAC scenario to emulate a 100m $\times$ 80m urban road intersection, featuring vehicles on the ground and low-altitude flying UAVs. Along the roadside, a BS is equipped with a UPA and three RGB-D cameras.
\vspace{-0.1cm}
\begin{itemize}
    \item \textbf{Scenario construction}: We use Simulation of Urban MObility (SUMO) software to simulate the trajectories of moving vehicles and UAVs at the intersection \cite{SUMO}, with the road width being 30 m. The vehicle trajectories include straight motion, turning and lane-changing, while the UAV trajectories encompass horizontal straight-line flight and vertical lift. The world coordinates of the vehicles and UAVs in each frame are exported as the positioning ground truth for the angle estimation and the off-line model training.
    \item \textbf{Visual data acquisition}: We construct a simulated urban-area scene in AirSim software by leveraging the real-physics engine \cite{airsim}. As the vehicles and UAVs are in motion, we use 3 cameras installed on the BS oriented towards the left, middle, and right perspectives, to capture RGB images and depth maps. This visual information serves as training data for the candidate object detector and ViR-Net. For the object detector, we annotate the bounding boxes to train it. Regarding ViR-Net, unsupervised training is carried out by using the acquired channel truth and position truth.
    \item \textbf{Channel data acquisition}: We imported the scene into the Wireless InSite® software \cite{WI} and configured the parameters of the BS and vehicle antenna arrays, along with the surface parameters, to ensure its consistency with the scene in AirSim. Post-scene import, communication and sensing channels for each frame are generated through ray tracing, in accordance with the channel model detailed in Section II. This channel information forms the basis for loss calculation and model training.
\end{itemize}

\vspace{-0.2cm}
\subsection{Parameters Setting}
Unless otherwise specified, the system setup for the hardware, the channels, the physical environment and the hyper-parameters in model training can be found in Table~\ref{tab:hardware}, Table~\ref{tab:channel}, Table~\ref{tab:visual} and Table~\ref{tab:model}, respectively. We define the SNR as $\frac{P_{t}}{Bn_{\text{c},0}}$. We conducted the training of the neural networks on an NVIDIA RTX4060Ti and tested the overall schemes in MATLAB 2024b. The system has 10 cores, a per-core clock speed of 2.5 GHz, and 160 floating-point operations per cycle.

\vspace{-0.25cm}
\subsection{Comparison Methods}
The methods to be evaluated include:
\vspace{-0.05cm}
\begin{itemize}
    \item \textbf{SoM-ISAC}: The proposed SoM-enhanced vision-RF integrated ISAC transmission strategy.
    \item \textbf{SA-Opt-ISAC}: A near-optimal benchmark with only RF operations. The BS first implements the initial hierarchical beam training as operated in \cite{beam_training_UPA} to acquire coarse-grained angular information, followed by the proposed SA-CP-BT in Section \RNum{3} and digital channel estimation to obtain the complete communication and sensing channel information as the prior. Then a squint-aware optimization of hybrid precoding is implemented as operated in \cite{FR} for ISAC transmission.
    The proposed RF-only squint-aware optimization scheme for ISAC.
    \item \textbf{Vi-ISAC}: An ablation approach for the proposed SoM-ISAC. The BS uses only visual information for end-to-end ISAC precoder design without RF operations.
    \item \textbf{BCD-ISAC}: An RF-only block coordinate descent (BCD)-based optimization approach without TTDs for ISAC in sub-THz systems \cite{ISAC_THz_opt}.
    \item \textbf{DPP}: Communication-centric squint elimination in \cite{DPP}.
    \item \textbf{YOLO}: Sensing-centric squint control in \cite{CBS}.
\end{itemize}

\vspace{-0.2cm}
\subsection{Evaluation Metric}
To comprehensively assess the performance of the sub-THz ISAC system, we take into account both the ISAC performance, SE and CRB, in data transmission, and the time overhead for acquiring the prior information and precoder design.
We define ISAC efficiency as the achievable time-averaged SE-CRB pair in each frame, i.e.,
\vspace{-0.1cm}
\begin{equation}
    \mathcal{E}_{\text{ISAC}}=(\overline{\mathcal{R}},\overline{\mathcal{CRB}}).
\vspace{-0.1cm}
\tag{25}
\end{equation}
The time-averaged SE and CRB are calculated respectively as
\vspace{-0.1cm}
\begin{equation}
    \overline{\mathcal{R}}=\mathcal{R}^{\star}\frac{T_{\text{data}}N_{\text{sub}}}{T_{\text{frame}}},~~~\overline{\mathcal{CRB}}=\mathcal{CRB}^{\star}\frac{T_{\text{frame}}}{T_{\text{data}}N_{\text{sub}}},
    \vspace{-0.0cm}
    \tag{26}
\end{equation}
where $\mathcal{R}^{\star}$ and $\mathcal{CRB}^{\star}$ denote the achievable SE-CRB pair on the ISAC performance boundary based on the corresponding scheme.  $T_{\text{frame}}$ is the duration per frame, $T_{\text{data}}$ is the duration for each data transmission stage. $T_{\text{frame}}=T_{\text{SSB}}+N_{\text{sub}}(T_{\text{RS}}+T_{\text{data}})$ for RF-only schemes, and $T_{\text{frame}}=N_{\text{sub}}(T_{\text{RS}}+T_{\text{data}})$ for vision-aided schemes, with $T_{\text{SSB}}$, $T_{\text{RS}}$ denoting the period of SSB and RS stages.

\begin{figure}[t]
  \vspace{-0.0cm}
  \centering
  \includegraphics[width=0.72\linewidth]{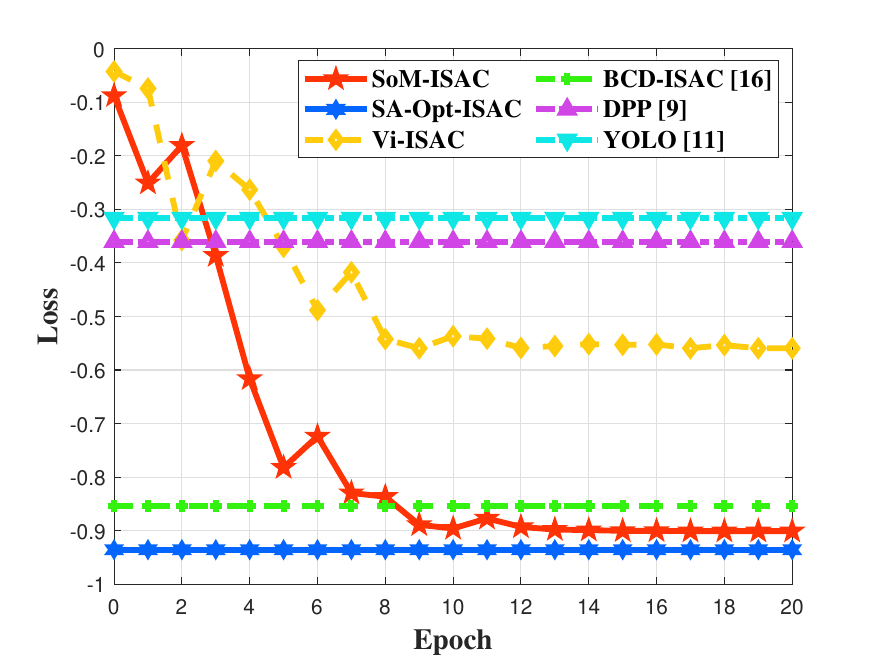}
  \vspace{-0.1cm}
  \captionsetup{font=small}
  \caption{Network training.}
  \label{fig:training}
  \vspace{-0.6cm}
\end{figure}

\begin{table}[b]
    \vspace{-0.4cm}
    \centering
    \captionsetup{font=small}
    \caption{Ablation Experiments in ViR-Net}
    \vspace{-0.1cm}
    \begin{tabular}{p{1.6cm} |p{1.1cm} |p{1.5cm} |p{1.2cm} |p{1.2cm}}
    \hline
         &  Loss $\downarrow$ & Correlation $\uparrow$ & SE $\uparrow$ & CRB $\downarrow$ \\
       \hline
       ViR-Net & \textbf{-0.906} & \textbf{0.9243}  & \textbf{31.23} &  \textbf{0.00294}    \\
       \hline
       w/o PE & -0.897 & 0.9067  & 29.34 & 0.00312     \\
       \hline
       w/o FFTF & -0.884 & 0.8988 & 29.32 & 0.00327     \\
       \hline
       SA-Opt-ISAC & \textbf{-0.928} & \textbf{0.9368} & \textbf{34.37} & \textbf{0.00258}     \\
       \hline
       BCD-ISAC & -0.854 & 0.6883 & 25.37 & 0.00328     \\
       \hline
    \end{tabular}
    \label{tab:ablation}
    \vspace{-0.0cm}
\end{table}

\vspace{-0.2cm}
\subsection{Comparative Study}
We first validate the performance of the proposed ViR-Net model during the training process in Fig. \ref{fig:training}. As the number of training epochs increases, the loss function of ViR-Net gradually decreases and rapidly converges after 10 epochs. When only visual information is input to ViR-Net, the model's loss increases significantly, implying the indispensable role of RF data. Ablation experiments in Table~\ref{tab:ablation} further confirm that the proposed approach improves SE, CRB, and channel correlation, while also demonstrating the positive impact of the Transformer layer and position encoding in the model.

\begin{figure*}[t]
\vspace{-0.0cm}
\centering
\subfigure[ISAC performance boundary.]{\label{fig:ISAC bound}
\includegraphics[width=0.33\linewidth]{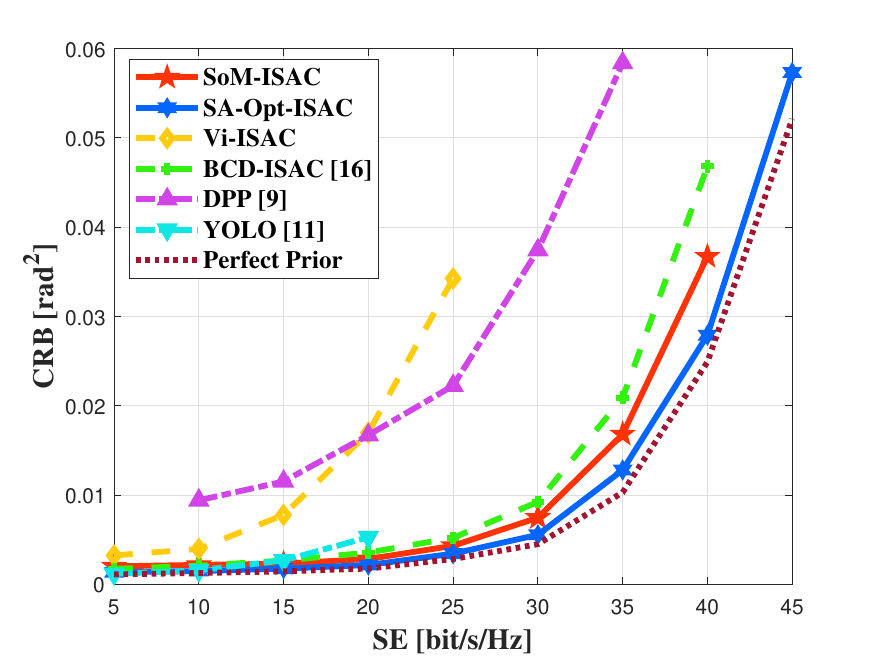}}
\hspace{-0.04\linewidth}
\subfigure[CRB versus MSIA.]{\label{fig:msia_crb}
\includegraphics[width=0.33\linewidth]{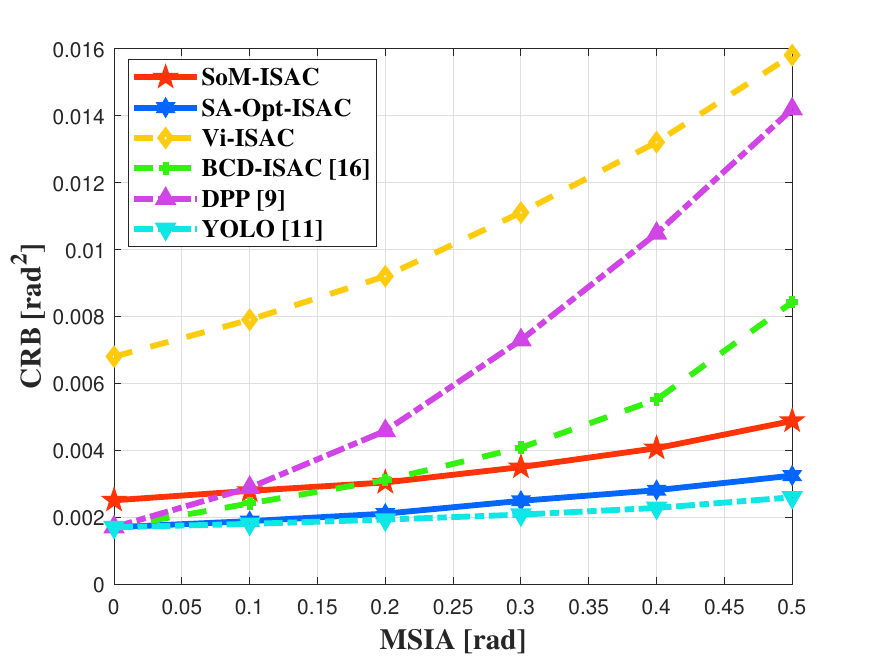}}
\hspace{-0.04\linewidth}
\subfigure[SE versus MSIA.]{\label{fig:msia_rate}
\includegraphics[width=0.33\linewidth]{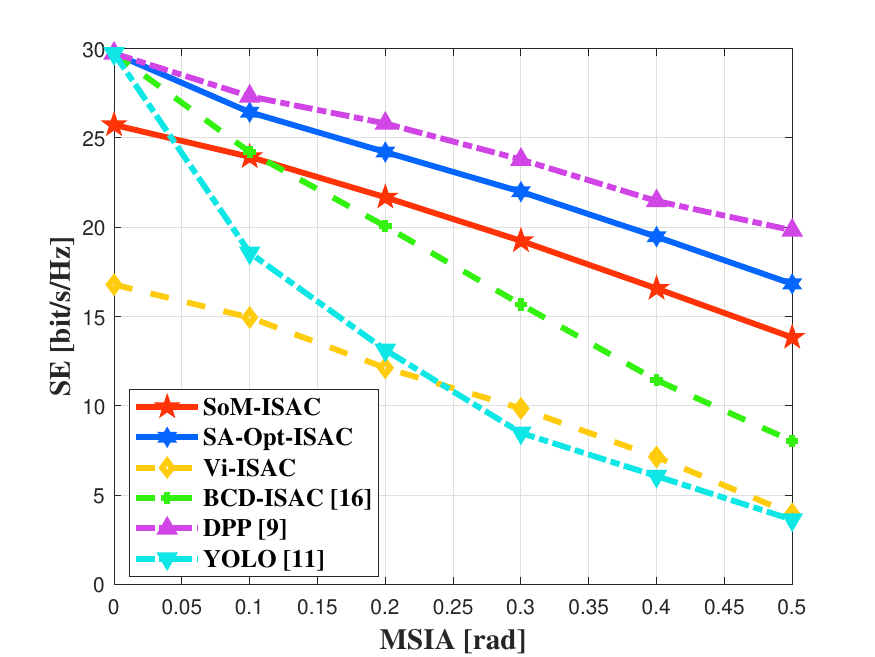}}
\vspace{-0.0cm}
\captionsetup{font=small}
\caption{ISAC performance comparison among different schemes.}
\label{fig:SNR}
\vspace{-0.3cm}
\end{figure*}

We compare the ISAC performance boundaries of different schemes in Fig.~\ref{fig:ISAC bound}, by optimizing the CRB with varying SE thresholds and SNR fixed at 10 dB. The SA-Opt-ISAC experiences errors in prior channel estimation, resulting in performance loss compared to the perfect prior case. However, it remains close to the theoretical optimal bound. The SoM-ISAC is inferior to SA-Opt-ISAC, as the visual information combined with reduced RF overhead does not provide more accurate prior data, but it still offers a better ISAC trade-off than the existing RF-only schemes. Compared to the non-TTD optimization scheme BCD-ISAC, TTD proves advantageous in actively regulating the correlation between communication and sensing channels, leading to better ISAC performance trade-offs. The Vi-ISAC, however, performs poorly, highlighting that relying solely on visual information in sub-THz systems without RF overheads is less effective.

\begin{figure}[t]
  \vspace{-0.0cm}
  \centering
  \includegraphics[width=0.74\linewidth]{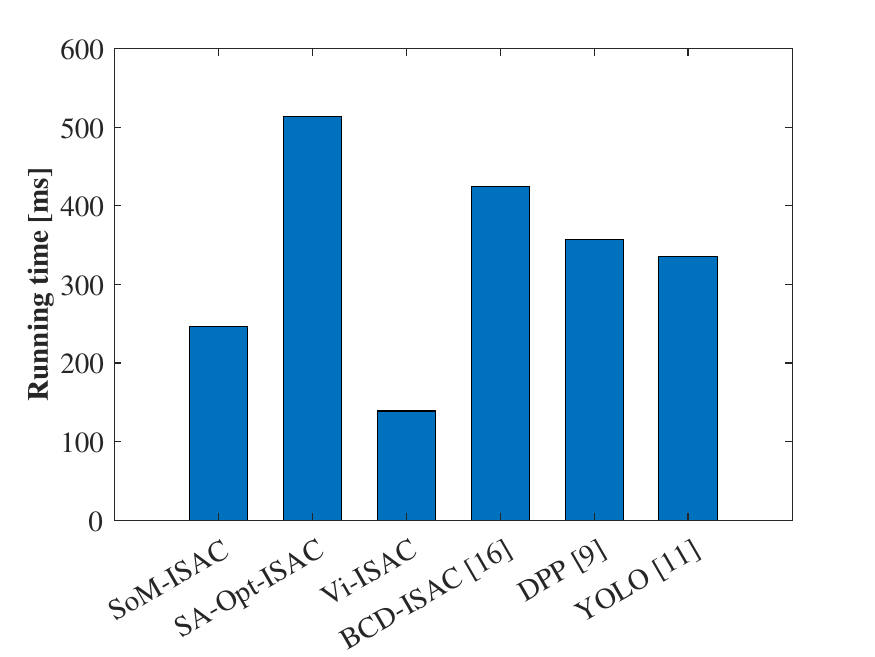}
  \vspace{-0.1cm}
  \captionsetup{font=small}
  \caption{Running time comparison among different precoding schemes.}
  \label{fig:time}
  \vspace{-0.2cm}
\end{figure}

\begin{figure}[t]
  \vspace{-0.0cm}
  \centering
  \includegraphics[width=0.74\linewidth]{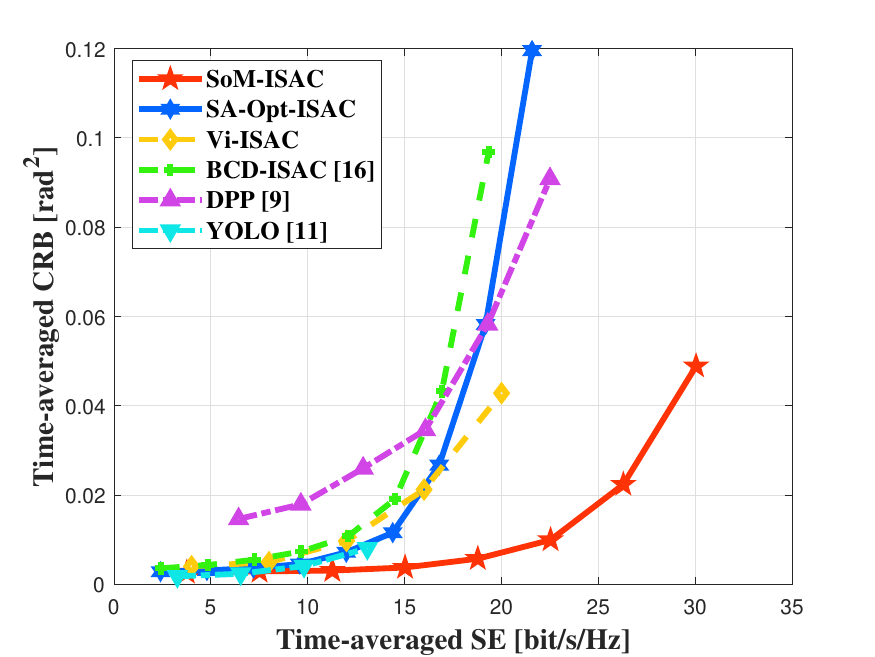}
  \vspace{-0.1cm}
  \captionsetup{font=small}
  \caption{Comparisons of ISAC efficiency among various schemes.}
  \label{fig:ISAC efficiency}
  \vspace{-0.4cm}
\end{figure}

Furthermore, we compare these schemes under different user-target spatial distributions in Fig.~\ref{fig:msia_crb} and Fig.~\ref{fig:msia_rate}, with SNR fixed at 10 dB. We define the mean squared interval of angle (MSIA) between targets and the user as $\sqrt{\frac{\sum_{u=1}^{U}\sum_{k=1}^{K}(\theta_{\text{s},k}-\theta_{\text{c},u})^{2}+(\phi_{\text{s},k}-\phi_{\text{c},u})^{2}}{2UK}}$. When MSIA=0 (indicating perfect user-target alignment), the RF-only schemes achieve optimal performance by focusing all beams. However, due to the limited dispersion of beam squint, the array gain decreases as the user-target angular separation increases, causing the ISAC performance of existing benchmarks to deteriorate significantly. Notably, the SA-Opt-ISAC and SoM-ISAC schemes effectively mitigate the performance loss by actively tuning the channels through TTDs, demonstrating better adaptation to spatial distribution.

In Fig.~\ref{fig:time}, we further compare the full running time of these schemes, incorporating both prior acquisition and precoder design.
The SA-Opt-ISAC has the highest running time due to the time-consuming initial beam training, channel estimation, and precoder design processes in each frame. In contrast, SoM-ISAC avoids these initial steps, with its primary time consumption coming from network inference, resulting in an overall runtime of approximately half that of SA-Opt-ISAC. The Vi-ISAC scheme, with zero RF overhead, has the least time overhead, which can be considered as an extreme case.

Taking the comprehensive operational time into account, we evaluate the ISAC efficiency boundaries of different schemes in Fig. \ref{fig:ISAC efficiency}, with a frame period of 1 s and 10 sub-frames per frame. Despite the performance gap in absolute terms, the proposed SoM-ISAC demonstrates significantly higher efficiency than the RF-only schemes at the frame level. This enhanced efficiency results from the balance between ISAC performance and time overhead, allowing more time for data transmission and target sensing within each frame. In contrast, the RF-only schemes exhibit lower efficiency, particularly at high SE, where its sensing efficiency drops significantly and even becomes inferior to that of the vision-only scheme. Therefore, it is evident that the proposed SoM-ISAC offers superior efficiency for real-world ISAC tasks.

\begin{figure*}[t]
\vspace{-0.0cm}
\centering
\captionsetup{font={small}}
\begin{minipage}[b]{.49\linewidth}
\centering
\includegraphics[scale=0.45]{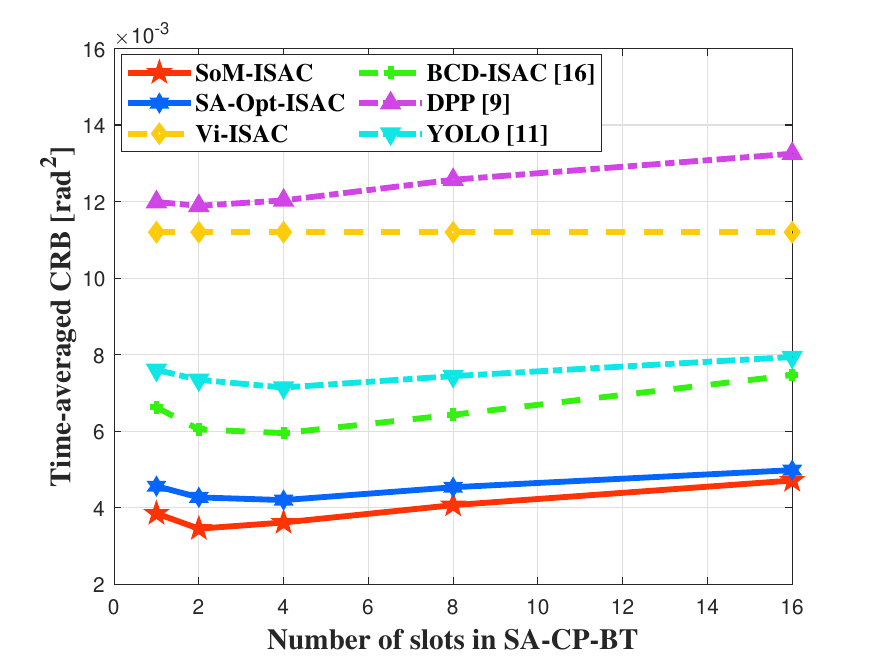}
\caption{Time-averaged CRB versus number of slots in SA-CP-BT.}
\label{fig:impact_steps}
\end{minipage}
\begin{minipage}[b]{.49\linewidth}
\centering
\includegraphics[scale=0.45]{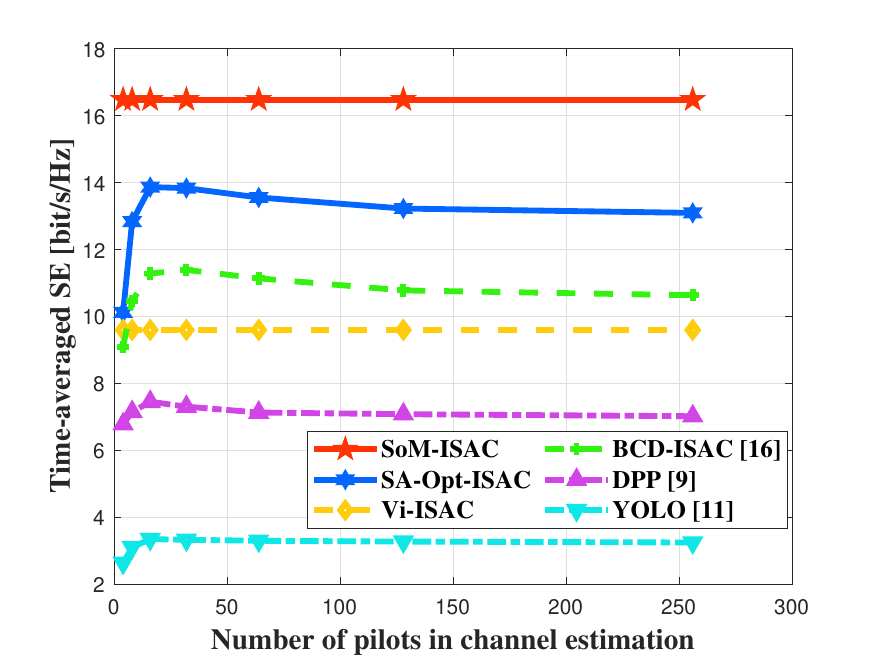}
\caption{Time-averaged SE versus number of pilots.}
\label{fig:pilots}
\end{minipage}
\begin{minipage}[b]{.49\linewidth}
\centering
\includegraphics[scale=0.45]{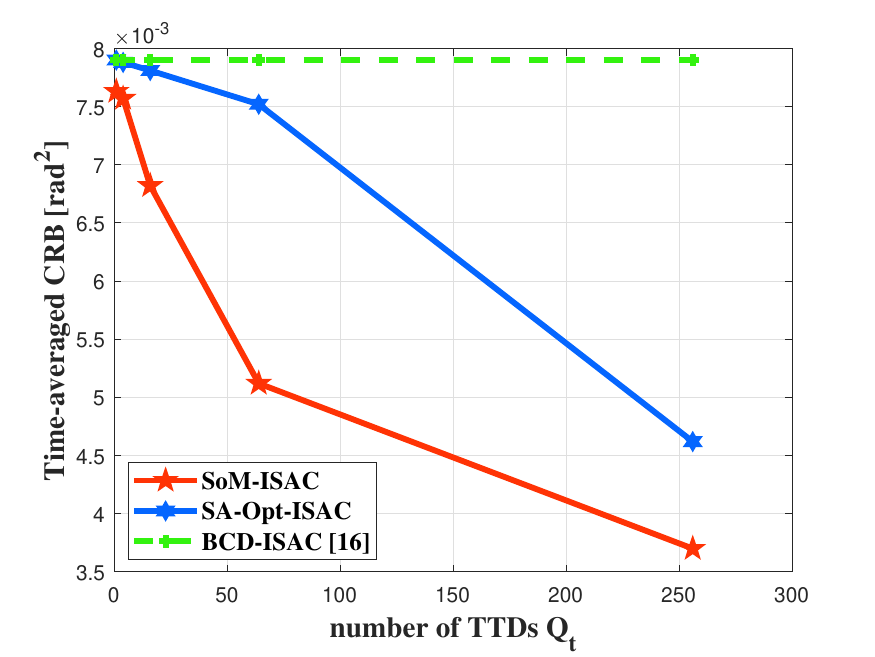}
\caption{Time-averaged CRB versus number of TTDs $Q_{t}$.}
\label{fig:crb_q}
\end{minipage}
\begin{minipage}[b]{.49\linewidth}
\centering
\includegraphics[scale=0.45]{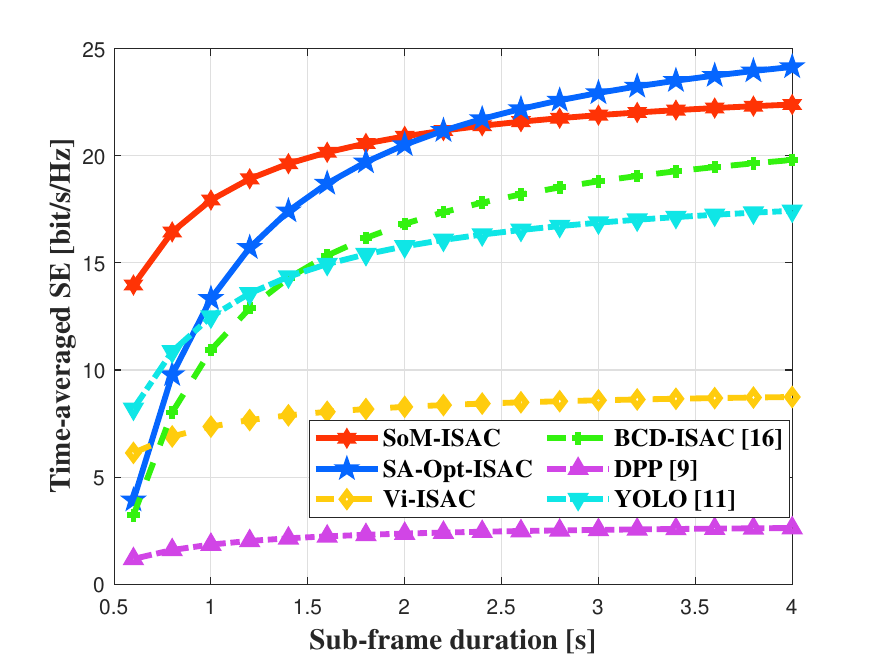}
\caption{Time-averaged SE versus subframe duration.}
\label{fig:frame_se}
\end{minipage}
\vspace{-0.25cm}
\end{figure*}

Additionally, we investigate the impact of system configurations on ISAC efficiency in following experiments. In Fig. \ref{fig:impact_steps}, we compare the influence of the number of slots in SA-CP-BT. A similar trend is observed in both RF-only and vision-aided schemes: as the number of slots increases, the time-averaged sensing efficiency initially improves but then declines. This is because increasing the number of slots improves the accuracy of prior angle estimation, which in turn enhances the precoding performance. However, once the number of slots exceeds a certain threshold, the performance gain no longer compensates for the increased time overhead, leading to a decrease in overall efficiency. The SoM-ISAC is more sensitive to the number of slots, as the time required for beam tracking constitutes a larger portion of the total operation time, resulting in greater variation. Nevertheless, the SoM-ISAC scheme still outperforms the RF-only schemes.

Subsequently, we examine the effect of the number of pilots on ISAC efficiency in Fig.~\ref{fig:pilots}. With different pilot configurations, the SA-Opt-ISAC scheme maintains its superiority in the RF-only category. All RF-only schemes show a trend of increasing and then decreasing efficiency as the number of pilots increases, indicating that a moderate number of pilots can enhance efficiency. In contrast, the vision-aided scheme operates with zero pilots, as it does not require channel estimation, yet its efficiency consistently surpasses that of RF-only schemes due to the complementary enhancement from visual information. This highlights the efficiency of the proposed vision-aided zero-pilot scheme.

In terms of hardware configurations, we compare ISAC efficiency under different TTD numbers in Fig.~\ref{fig:crb_q}. When $Q_{\text{t}}=N_{\text{t}}$, the TTD offers maximum flexibility, enabling the achievement of the optimal CRB and SE for the highest efficiency. As the number of TTDs decreases, both the accuracy of beam tracking and the DoF in correlation regulation diminish, leading to performance degradation and subsequent reduction in ISAC efficiency. Thus, the role of TTDs in enhancing ISAC efficiency in sub-THz systems is demonstrated. This trend is more pronounced in the SA-Opt-ISAC, where the reduction in the number of TTDs amplifies the errors in channel prior information. In contrast, the SoM-ISAC is relatively less affected by the reduction in accuracy.

Finally, we compare the ISAC efficiency across different frame periods in Fig.~\ref{fig:frame_se}. In real-world scenarios, varying frame periods correspond to different user and target speeds. As the speeds of users and targets increase, their positions change more rapidly, requiring more frequent beam tracking, which results in a shorter subframe period. In the experiments, as the subframe period increases, the communication and sensing efficiency of each scheme gradually improves until it stabilizes, reaching a convergence point in a static scenario. Notably, the efficiency of the SA-Opt-ISAC improves more significantly with longer subframe periods, eventually approaching or even surpassing that of the SoM-ISAC when the subframe period becomes sufficiently long. This indicates that SA-Opt-ISAC is better suited for quasi-static scenarios with longer subframes, while the SoM-ISAC is more advantageous for dynamic scenarios with shorter subframes.

\vspace{-0.25cm}
\section{Conclusions}

This paper presented a SoM-enhanced framework for sub-THz ISAC transmission in air-ground networks. By leveraging the design degrees of freedom in sub-THz hardware and channels, and by integrating multi-modal sensing information, the proposed solution significantly improved ISAC performance while accommodating three-dimensional air-ground dynamics.
The approach began with the utilization of visual information and sparse RF operations to swiftly acquire essential prior knowledge about mobile users and targets within three-dimensional networks. A multi-modal fusion learning technique was then developed to optimize the ISAC hybrid precoder, achieving a better balance between dual functionalities.
Comprehensive simulations demonstrated that the proposed framework not only ensured robust ISAC performance in the sub-THz range but also minimized operational time overhead, leading to a substantial enhancement in ISAC efficiency.

\vspace{-0.2cm}
\appendices

\section{Proof of the Proposition 1}
According to \cite{IT_ISAC_Liufan}, the optimal $\bm R_{x,m}$ can be expressed as 
\vspace{-0.2cm}
\begin{equation}
\bm R_{x,m}^{\star}=\bm U_{m}\bm \Lambda_{m}\bm U_{m}^{\text{H}}, \nonumber
\vspace{-0.2cm}
\end{equation}
where $\bm U_{m}\!=\![\bm A_{t,m}^{*}, \dot{\bm A}_{t,\theta,m}^{*},\dot{\bm A}_{t,\phi,m}^{*}, \{\bm a_{t}(\theta_{\text{c},u},\phi_{\text{c},u},f_{m})\}_{u=1}^{U}]\!=\![\bm u_{m,1}, \cdots, \!\bm u_{m,N_{R}}]\!\in \!\mathbb{C}^{N_{t}\times N_{R}}$ ($N_{R}\!=\!3K\!+\!U$) serving as the basis vectors under the given channels $\bm H_{m}$ and $\bm G_{m}$, and $\bm \Lambda_{m}$ is positive semi-definite with $\bm \Lambda_{m}[n_1,n_2]=\lambda_{m,n_{1},n_{2}}$. 
Then the achievable rate-CRB on the Pareto boundary is written as
\vspace{-0.2cm}
\begin{align}
    &\mathcal{R}^{\star}\!=\!\!\sum_{m=1}^{M}\log_{2}\!\left[1\!+\!\frac{1}{\sigma_{c}^{2}}\sum_{n_{1}=1}^{N_{R}}\sum_{n_{2}=1}^{N_{R}}\lambda_{m,n_{1},n_{2}}\bm \Xi_{\text{c},m}[n_{1},n_{2}]\right],\nonumber \\
    \vspace{-0.3cm}
    &\mathcal{CRB}^{\star}\!\approx\! N_{R}^{2}\!\left [\sum_{m=1}^{M}\sum_{n_{1}=1}^{N_{R}}\sum_{n_{2}=1}^{N_{R}}\!\lambda_{m,n_{1},n_{2}} \bm \Xi_{\text{s},m}[n_{1},n_{2}]\!\right ]^{-1}\!,
    \vspace{-0.5cm}
    \label{equ:optimal}
    \tag{27}
\end{align}
$\bm \Xi_{\text{c},m}[n_{1}\!,\!n_{2}]\!\!=\!\!\xi_{\text{c},m,n_{1},n_{2}}\!\!\!=\!{\bm h_{\text{c},m}^{\text{b}}}^{\text{H}}\bm D_{t}^{\text{H}}\bm u_{m,n_{1}}\bm u_{m,n_{2}}^{\text{H}}\bm D_{t}\bm h_{\text{c},m}^{\text{b}}$, $\bm \Xi_{\text{s},m}[n_{1}\!,\!n_{2}]\!\!=\!\!\xi_{\text{s},m,n_{1},n_{2}}\!\!\!=\!\bm u_{m,n_{1}}^{\text{H}}\bm A_{t,m}\bm \Sigma_{m}\bm \Sigma_{m}^{*}\bm A_{t,m}^{\text{H}}\bm u_{m,n_{2}}$ are both determined by the channels. By solving this multi-objective optimization, the optimal $\bm \Lambda_{m}$ can be written as
\vspace{-0.2cm}
\begin{equation}
    \bm \Lambda_{m}^{\star}=( \bm \Xi_{\text{c},m}+\gamma \bm \Xi_{\text{s},m})\frac{P_{t}}{M\text{Tr}(\bm \Xi_{\text{c},m}+\gamma \bm \Xi_{\text{s},m})}.
    \tag{28}
    \label{equ:Lambda}
    \vspace{-0.15cm}
\end{equation}
By substituting (\ref{equ:Lambda}) into (\ref{equ:optimal}), the rate and CRB at the Pareto optimum can be described via $\bm \Xi_{\text{c},m}$ and $\bm \Xi_{\text{s},m}$ as
\vspace{-0.1cm}
\begin{equation}
\begin{split}
    \!\!\!\!\mathcal{R}^{\star}\!\!\!=&\!\!\sum_{m}^{M}\!\log_{2}\!\!\left[\!1\!\!
    + \!\frac{1}{\sigma_{\text{c}}^{2}}\!\!\sum_{n_{1}}^{N_{R}}\!\!\sum_{n_{2}}^{N_{R}}\!\!\frac{P_{t}(\xi_{\text{c},m,n_{1},n_{2}}^{2}\!\!+\!\!\gamma \xi_{\text{s},m,n_{1},n_{2}}\xi_{\text{c},m,n_{1},n_{2}})}{M\text{Tr}(\bm \Xi_{\text{c},m}\!+\!\gamma \bm \Xi_{\text{s},m})}\!\!\right]\!\!,
\end{split}
\tag{29}
\end{equation}
\begin{equation}
\begin{split}
\!\!\!\!&\mathcal{CRB}^{\star}\!\!\approx\!\!N_{R}^{2}\!\!\left[\!\sum_{m}^{M}\!\sum_{n_{1}}^{N_{R}}\!\sum_{n_{2}}^{N_{R}}\!\frac{P_{t}(\gamma \xi_{\text{s},m,n_{1},n_{2}}^{2}\!\!\!\!+\!\!\xi_{\text{s},m,n_{1},n_{2}}\xi_{\text{c},m,n_{1},n_{2}})}{M\text{Tr}(\bm \Xi_{\text{c},m}\!+\!\gamma \bm \Xi_{\text{s},m})}\!\!\right]^{-1}\!\!\!\!. 
\end{split}
\tag{30}
\end{equation}
Leveraging the sparsity of beamspace channels \cite{shijian_MI_max}, we denote the indices of peaks in $\widehat{\bm h}_{\text{c}}^{\text{b}}$ and $\widehat{\bm h}_{\text{s}}^{\text{b}}$ as $\{\psi_{\text{c},m}\}$ and $\{\psi_{\text{s},k,m}\}$. The similarity of these peaks is defined as $\mathcal{S}_{m}(\bm H,\bm G)=\sum_{k=1}^{K}\mathbb{I}( \psi_{\text{c},m}=\psi_{\text{s},k,m})$ and $\mathcal{S}(\bm H,\bm G)=\sum_{m=1}^{M}\mathcal{S}_{m}$ with $\mathbb{I}(\cdot)$ being the indicator function. It can be proved that a higher $\text{Cor}(\bm H,\bm G)$ gives rise to a higher $\mathcal{S}(\bm H,\bm G)$ \cite{KL_proof}.
As derived in Eq.~(\ref{equ:xi}), as $\mathcal{S}(\bm H,\bm G)$ increases, both $\xi_{\text{c},m,n_{1},n_{2}}$ and $\xi_{\text{s},m,n_{1},n_{2}}$ improve, leading to an increase in $\mathcal{R}^{\star}$ and a reduction in $\mathcal{CRB}^{\star}$ along the Pareto boundary. Therefore a higher C-S channel correlation improves the dual-functional gain.

\begin{figure*}[t]
\vspace{-0.35cm}

     \begin{align}
    \xi_{\text{c},m,n_{1},n_{2}}\!&\!\approx \!\mathbb{I}(n_{1}\!\!=\!n_{2}\!\!>\!\!3K)\!\!+\!\!2\mathbb{I}(1\!\!\leqslant \!\!n_{1}\!\!\leqslant\!\! K,\!n_{2}\!\!>\!\!3K)\!\!\sum_{k=1}^{K}\!\mathbb{I}(\psi_{\text{s},k,m}\!\!=\!\!\psi_{\text{c},m})\!\!+\!\! \mathbb{I}(1\!\!\leqslant \!\!n_{1},\!n_{2}\!\!\leqslant\!\! K)\sum_{k_{1}}^{K}\!\!\sum_{k_{2}}^{K}\!\mathbb{I}(\psi_{\text{s},k_{1},m}\!\!=\!\!\psi_{\text{c},m})\mathbb{I}(\psi_{\text{s},k_{2},m}\!\!=\!\!\psi_{\text{c},m}),\nonumber  \\
    \vspace{-0.99cm}
    \xi_{\text{s},m,n_{1},n_{2}}\!&\!\approx \!\mathbb{I}(n_{1}\!=\!n_{2}\!>\!3K)\sum_{k=1}^{K}\mathbb{I}(\psi_{\text{c},m}\!\!=\!\psi_{\text{s},k,m})\vert\bm \Sigma_{m}[k,k]\vert^{2} \!+\!\mathbb{I}(1\leqslant n_{1},\!n_{2}\!\leqslant\!K)\sum_{k=1}^{K}\vert\bm \Sigma_{m}[k,k]\vert^{2}.
    \label{equ:xi}
    \vspace{-0.2cm}
    \tag{31}
\end{align}
\hrulefill
\vspace{-0.35cm}
\end{figure*}

\section{Proof of the Proposition 2}
\vspace{-0.2cm}
Denote the array gain towards the direction $(\phi, \theta)$ at the $m$-th subcarrier, with the TTDs being $\bm T$ and the PSs connecting to one RF chain being $\bm \varphi$, as 
\vspace{-0.2cm}
\begin{align}
   \!\!\!\!&g( \theta, \phi, f_{m}; \bm T, \bm \varphi)=\bm a^{\text{H}}(\theta, \phi)\bm F_{\text{T},m}\bm F_{\text{P}} \nonumber \\
    \!\!\!\!&\!=\!\!\!\biggl|\!\sum_{n_{t_{h}}}\!\!\sum_{n_{t_{v}}}\!\!e^{j\pi\left(\!\bm \varphi[n_{t_{h}}\!,n_{t_{v}}]\!-\!2f_{m}\bm T[n_{t_{h}}\!,\!n_{t_{v}}]\!-\!\frac{f_{m}}{f_{\text{c}}}[\sin\!\theta\sin\!\phi(n_{t_{h}}\!-\!1)\!+\cos\!\theta(n_{t_{v}}\!-\!1)] \!\right)} \biggr|.  \nonumber
    \vspace{-0.25cm}
\end{align}

Note that $g(\theta, \phi, f_{m}; \bm T, \bm \varphi)$ reaches its maximum if and only if each term of $g( \theta, \phi, f_{m}; \bm T, \bm \varphi)$ has phase 0 for any $(n_{t_{h}},n_{t_{v}})$. When $Q_{\text{t}}=N_{\text{t}}$, to make the diverged beams in different subcarriers cover the angular trajectory from $(\phi_{0},\theta_{0})$ to $(\phi_{1},\theta_{1})$, it is necessary to make the beam at the first subcarrier point to the starting angle and the beam at the $M$-th subcarrier point to the ending angle, i.e.,
\vspace{-0.15cm}
\begin{align}
    \bm \varphi[n_{t_{h}},n_{t_{v}}]\!&-\!2f_{1}\bm T[n_{t_{h}},n_{t_{v}}]\!-\!\!\frac{f_{1}}{f_{\text{c}}}\biggl(\sin\theta_{0}\sin\phi_{0}(n_{t_{h}}\!-\!1)\! \nonumber\\ 
    &~~~~~~+\!\cos\theta_{0}(n_{t_{v}}\!-\!1)\biggr)\!=\!0, \nonumber \\
    \bm \varphi[n_{t_{h}},n_{t_{v}}]\!&-\!2f_{M}\bm T[n_{t_{h}},n_{t_{v}}]\!-\!\!\frac{f_{M}}{f_{\text{c}}}\biggl(\sin\theta_{1}\sin\phi_{1}(n_{t_{h}}-1) \nonumber \\ 
    &~~~~~~+\!\cos\theta_{1}(n_{t_{v}}\!-\!1)\biggr)\!=\!0. \nonumber
\end{align}
Solving this set of equations gives rise to (\ref{equ:CBS_TTD}) and (\ref{equ:CBS_PS}). By substituting  (\ref{equ:CBS_TTD}) and (\ref{equ:CBS_PS}) into $g(\theta, \phi, f_{m}; \bm T, \bm \varphi)$, we have the beam at the $m$-th subcarrier point to the direction
\vspace{-0.2cm}
\begin{align}
    \phi_{m}\!\!&=\!\arcsin\! \left( \frac{(f_{M}\!\!-\!\!f_{m})f_{1}}{Bf_{m}}\!\sin \phi_{0}\!\!-\!\frac{f_M(f_{1}\!\!-\!\!f_{m})}{f_{m}B}  \sin \phi_{1} \!\right ),\nonumber\\
    \theta_{m}\!&=\arccos \left( \frac{(f_{M}-f_{m})f_{1}}{Bf_{m}}\cos \theta_{0}-\frac{f_M}{f_{m}}  \cos \theta_{1}  \right ). \nonumber
\end{align}
It is easy to prove that both $\phi_{m}$ and $\theta_{m}$ are monotonous with the index. When $Q_{\text{t}}\leqslant N_{\text{t}}$, simply averaging the delay values within the same subarray can achieve the beam pointing control, giving rise to Eq.~(\ref{equ:CBS_TTD}) and Eq.~(\ref{equ:CBS_PS}), and the above conclusion still holds true.

Controlling the beam squint in the angular trajectory from $(\phi_{0},\theta_{0})$ to $(\phi_{1},\theta_{1})$, the array gain at the $m$-th subcarrier received by the user within this range can be expressed as
\vspace{-0.2cm}
\begin{align}
    &g(\theta_{c,u}, \phi_{c,u}, f_{m}; \bm T, \bm \varphi)=\bm a^{\text{H}}(\theta_{c,u}, \phi_{c,u}, f_{m})\bm a_{\theta_{m},\phi_{m},f_{m}} \nonumber \\
    &\!\!=\!\!\biggl(\!\bm a_{h}^{\text{H}}(\theta_{c,u}, \!\phi_{c,u},\! f_{m}\!)\bm a_{h}\!(\theta_{m},\!\phi_{m},\!f_{m})\biggr) \!\biggl(\bm a_{v}^{\text{H}}(\!\theta_{c,\!u},\! f_{m})\bm a_{v}\!(\theta_{m},f_{m}\!)\!\biggr) \nonumber\\
    &=\frac{1}{N_{t}^{2}}\biggl| \Phi\left(\frac{f_{m}}{f_{\text{c}}}(\sin\phi_{m}\sin\theta_{m}-\sin\phi_{c,u}\sin\theta_{c,u}) \right)\biggr|  \nonumber\\
    &~~~\cdot \biggl| \Phi\left(\frac{f_{m}}{f_{\text{c}}}(\cos\theta_{m}-\cos\theta_{c,u})\right)\biggr|, \nonumber
\end{align}
where $\Phi(x)=\frac{\sin(N_{t}\pi x/2)}{\sin(\pi x/2)}$ is the Dirichlet function. Note that the main lobe of $\Phi(x)$ lies in the range $[-\frac{1}{N_t}, \frac{1}{N_t}]$. 

When the angular constraints in the Proposition 2 are satisfied, we have
\vspace{-0.2cm}
\begin{equation}
\begin{aligned}
    &\biggl|\frac{f_{m}}{f_{\text{c}}}(\sin\phi_{m}\sin\theta_{m}-\sin\phi_{c,u}\sin\theta_{c,u})\biggr| \\
    &\leqslant  \frac{f_{M}}{f_{c}}\vert \sin\phi_{c,u}\vert\cdot \vert\theta_{m}-\theta_{c,u}\vert  \leqslant \frac{2}{N_{t_{h}}}, \\
    &\biggl|\frac{f_{m}}{f_{\text{c}}}(\cos\theta_{m}-\cos\theta_{c,u})\biggr| \\
    &\leqslant \frac{f_{M}}{f_{c}} \vert\phi_{m}-\phi_{c,u}\vert \leqslant \frac{2}{N_{t_{v}}}. \nonumber
\end{aligned}
\vspace{-0.1cm}
\end{equation}
Therefore the angular trajectory by the beam squint lies in the main lobe of $g(\theta_{c,u}, \phi_{c,u}, f_{m}; \bm T, \bm \varphi)$, and the monotonicity in the Proposition 2 is satisfied.

\ifCLASSOPTIONcaptionsoff
  \newpage
\fi

\vspace{-0.1cm}
\bibliographystyle{IEEEtran}
\bibliography{IEEEabrv,myrefs}

\end{document}